%From no-reply@arXiv.org Sun Sep 14 16:02:58 2003
%Date: Sun, 14 Sep 2003 08:33:44 -0400
%From: send mail ONLY to hep-th <no-reply@arXiv.org>
%Reply-To: hep-th@arXiv.org
%To: giveon@phys.huji.ac.il
%Subject: RE: uget 0309056

\input harvmac   
%\draftmode
\def\journal#1&#2(#3){\unskip, \sl #1\ \bf #2 \rm(19#3) }
\def\andjournal#1&#2(#3){\sl #1~\bf #2 \rm (19#3) }

\def\frac#1#2{{#1\over#2}}

\def\d{\partial}

\def\inbar{\,\vrule height1.5ex width.4pt depth0pt}
\def\IC{\relax\hbox{$\inbar\kern-.3em{\rm C}$}}
\def\IR{\relax{\rm I\kern-.18em R}}
\def\IP{\relax{\rm I\kern-.18em P}}
\def\IZ{\relax{\rm I\kern-.18em Z}}

%
%%%%%%%%%%%%%%%%%%%%%%%%%%%%%%%%%%%%
%

%
\catcode`\@=11
\def\slash#1{\mathord{\mathpalette\c@ncel{#1}}}
\overfullrule=0pt

\def\GG{\Gamma}

\def\LL{{\cal L}}

\def\SS{{\cal S}}

\def\eps{\epsilon}

\def\underrel#1\over#2{\mathrel{\mathop{\kern\z@#1}\limits_{#2}}}

\catcode`\@=12

%%%%%%%%%%%%%%%%%%%%%%%%%%%%%%%%%%%%%%%%%%%%%%%%%%%%%%%%%%%%%%

%

\def \sinh{{\rm sinh}}
\def \cosh{{\rm cosh}}

%%%%%%%%%%%%%%%%%%%%%%%%%%%%%%%%%%%%%%%%%%%%%%%%%%%%%%%%%%%%%%
% new defs:

\def\S{{\bf S}}
\def\T{{\bf T}}

\def\ed{\frac{\eps_2}{2}}

\def\e{\epsilon}
\def\k{\kappa}

\rightline{RI-08-03}
\Title{
\rightline{hep-th/0309056}}
{\vbox{\centerline{Type 0 Strings in a 2-d Black Hole}}}
\medskip
\centerline{\it Amit Giveon, Anatoly Konechny, Ari Pakman, Amit Sever}
\bigskip
\centerline{Racah Institute of Physics, The Hebrew University}
\centerline{Jerusalem 91904, Israel}
\centerline{giveon@vms.huji.ac.il, tolya@phys.huji.ac.il, 
pakman@phys.huji.ac.il, asever@phys.huji.ac.il}

\bigskip\bigskip\bigskip
\noindent
We study some aspects of 
type 0 strings propagating in the two dimensional black hole
geometry, corresponding to the exact $SL(2)/U(1)$ SCFT background.

\vfill

\Date{08/03}

\newsec{Introduction}

\lref\gk{
A.~Giveon and D.~Kutasov,
``Little string theory in a double scaling limit,''
JHEP {\bf 9910}, 034 (1999)
[arXiv:hep-th/9909110].
%%CITATION = HEP-TH 9909110;%%
}
\lref\gktwo{
A.~Giveon and D.~Kutasov,
``Comments on double scaled little string theory,''
JHEP {\bf 0001}, 023 (2000)
[arXiv:hep-th/9911039].
%%CITATION = HEP-TH 9911039;%%
}
\lref\hk{
K.~Hori and A.~Kapustin,
``Duality of the fermionic 2d black hole and 
N = 2 Liouville theory as  mirror symmetry,''
JHEP {\bf 0108}, 045 (2001)
[arXiv:hep-th/0104202].
%%CITATION = HEP-TH 0104202;%%
}
\lref\murthy{
S.~Murthy,
``Notes on non-critical superstrings in various dimensions,''
arXiv:hep-th/0305197.
%%CITATION = HEP-TH 0305197;%%
}
\lref\gkthree{
A.~Giveon and D.~Kutasov,
``Notes on AdS(3),''
Nucl.\ Phys.\ B {\bf 621}, 303 (2002)
[arXiv:hep-th/0106004].
%%CITATION = HEP-TH 0106004;%%
}
\lref\giv{
A.~Giveon,
``Target space duality and stringy black holes,''
Mod.\ Phys.\ Lett.\ A {\bf 6}, 2843 (1991).
%%CITATION = MPLAE,A6,2843;%%
}
\lref\dvv{
R.~Dijkgraaf, H.~Verlinde and E.~Verlinde,
``String propagation in a black hole geometry,''
Nucl.\ Phys.\ B {\bf 371}, 269 (1992).
%%CITATION = NUPHA,B371,269;%%
}
\lref\grs{
A.~Giveon, E.~Rabinovici and A.~Sever,
``Beyond the singularity of the 2-D charged black hole,''
JHEP {\bf 0307}, 055 (2003)
[arXiv:hep-th/0305140].
%%CITATION = HEP-TH 0305140;%%
}
\lref\six{
M.~R.~Douglas, I.~R.~Klebanov, D.~Kutasov, J.~Maldacena, 
E.~Martinec and N.~Seiberg,
``A new hat for the c = 1 matrix model,''
arXiv:hep-th/0307195.
%%CITATION = HEP-TH 0307195;%%
}
\lref\tt{
T.~Takayanagi and N.~Toumbas,
``A matrix model dual of type 0B string theory in two dimensions,''
JHEP {\bf 0307}, 064 (2003)
[arXiv:hep-th/0307083].
%%CITATION = HEP-TH 0307083;%%
}
\lref\kms{
I.~R.~Klebanov, J.~Maldacena and N.~Seiberg,
``D-brane decay in two-dimensional string theory,''
JHEP {\bf 0307}, 045 (2003)
[arXiv:hep-th/0305159].
%%CITATION = HEP-TH 0305159;%%
}
\lref\mv{
J.~McGreevy and H.~Verlinde,
``Strings from tachyons: The c = 1 matrix reloated,''
arXiv:hep-th/0304224.
%%CITATION = HEP-TH 0304224;%%
}
\lref\mtv{
J.~McGreevy, J.~Teschner and H.~Verlinde,
%``Classical and quantum D-branes in 2D string theory,''
arXiv:hep-th/0305194.
%%CITATION = HEP-TH 0305194;%%
}
\lref\zz{
A.~B.~Zamolodchikov and A.~B.~Zamolodchikov,
``Liouville field theory on a pseudosphere,''
arXiv:hep-th/0101152.
%%CITATION = HEP-TH 0101152;%%
}
\lref\gks{
A.~Giveon, D.~Kutasov and A.~Schwimmer,
``Comments on D-branes in AdS(3),''
Nucl.\ Phys.\ B {\bf 615}, 133 (2001)
[arXiv:hep-th/0106005].
%%CITATION = HEP-TH 0106005;%%
}
\lref\pst{
B.~Ponsot, V.~Schomerus and J.~Teschner,
``Branes in the Euclidean AdS(3),''
JHEP {\bf 0202}, 016 (2002)
[arXiv:hep-th/0112198].
%%CITATION = HEP-TH 0112198;%%
}
\lref\klebanov{
I.~R.~Klebanov,
``String theory in two-dimensions,''
arXiv:hep-th/9108019.
%%CITATION = HEP-TH 9108019;%%
}
\lref\gm{
P.~Ginsparg and G.~W.~Moore,
``Lectures On 2-D Gravity And 2-D String Theory,''
arXiv:hep-th/9304011.
%%CITATION = HEP-TH 9304011;%%
}
\lref\sen{
A.~Sen,
``Rolling tachyon,''
JHEP {\bf 0204}, 048 (2002)
[arXiv:hep-th/0203211];
``Tachyon matter,''
JHEP {\bf 0207}, 065 (2002)
[arXiv:hep-th/0203265];
``Field theory of tachyon matter,''
Mod.\ Phys.\ Lett.\ A {\bf 17}, 1797 (2002)
[arXiv:hep-th/0204143].
}
\lref\dfk{
P.~Di Francesco and D.~Kutasov,
``World sheet and space-time physics in two-dimensional 
(Super)string theory,''
Nucl.\ Phys.\ B {\bf 375}, 119 (1992)
[arXiv:hep-th/9109005].
%%CITATION = HEP-TH 9109005;%%
}
\lref\kkk{
V.~Kazakov, I.~K.~Kostov and D.~Kutasov,
``A matrix model for the two-dimensional black hole,''
Nucl.\ Phys.\ B {\bf 622}, 141 (2002)
[arXiv:hep-th/0101011].
%%CITATION = HEP-TH 0101011;%%
}
\lref\pol{J.~Polchinski,
``String Theory. Vol. 2: Superstring Theory And Beyond,''
Cambridge university press (1998).
}
\lref\efr{
S.~Elitzur, A.~Forge and E.~Rabinovici,
``Some Global Aspects Of String Compactifications,''
Nucl.\ Phys.\ B {\bf 359}, 581 (1991).
%%CITATION = NUPHA,B359,581;%%
}
\lref\witten{
E.~Witten,
``On string theory and black holes,''
Phys.\ Rev.\ D {\bf 44}, 314 (1991).
%%CITATION = PHRVA,D44,314;%%
}
\lref\kazsuz{
Y.~Kazama and H.~Suzuki,
``New N=2 Superconformal Field Theories And Superstring Compactification,''
Nucl.\ Phys.\ B {\bf 321}, 232 (1989);
``Characterization Of N=2 Superconformal 
Models Generated By Coset Space Method,''
Phys.\ Lett.\ B {\bf 216}, 112 (1989).
}
\lref\gpr{
A.~Giveon, M.~Porrati and E.~Rabinovici,
``Target space duality in string theory,''
Phys.\ Rept.\  {\bf 244}, 77 (1994)
[arXiv:hep-th/9401139].
%%CITATION = HEP-TH 9401139;%%
}
\lref\egkr{
S.~Elitzur, A.~Giveon, D.~Kutasov and E.~Rabinovici,
``From big bang to big crunch and beyond,''
JHEP {\bf 0206}, 017 (2002)
[arXiv:hep-th/0204189].
%%CITATION = HEP-TH 0204189;%%
}
\lref\tong{
D.~Tong,
``Mirror mirror on the wall: 
On two-dimensional black holes and Liouville  theory,''
JHEP {\bf 0304}, 031 (2003)
[arXiv:hep-th/0303151].
%%CITATION = HEP-TH 0303151;%%
}
\lref\gkrev{
A.~Giveon and D.~Kutasov,
``Brane dynamics and gauge theory,''
Rev.\ Mod.\ Phys.\  {\bf 71}, 983 (1999)
[arXiv:hep-th/9802067].
%%CITATION = HEP-TH 9802067;%%
}
\lref\kir{
E.~B.~Kiritsis,
``Duality in gauged WZW models,''
Mod.\ Phys.\ Lett.\ A {\bf 6}, 2871 (1991).
%%CITATION = MPLAE,A6,2871;%%
}
\lref\mgmv{
J.~McGreevy, S.~Murthy and H.~Verlinde,
``Two-dimensional superstrings and the supersymmetric matrix model,''
arXiv:hep-th/0308105.
%%CITATION = HEP-TH 0308105;%%
}
\lref\daps{
S.~P.~de Alwis, J.~Polchinski and R.~Schimmrigk,
``Heterotic Strings With Tree Level Cosmological Constant,''
Phys.\ Lett.\ B {\bf 218}, 449 (1989).
%%CITATION = PHLTA,B218,449;%%
}
\lref\kuse{
D.~Kutasov and N.~Seiberg,
``Noncritical Superstrings,''
Phys.\ Lett.\ B {\bf 251}, 67 (1990).
%%CITATION = PHLTA,B251,67;%%
}
\lref\pak{
A.~Pakman,
``Unitarity of supersymmetric SL(2,R)/U(1) and no-ghost theorem for  
fermionic strings in AdS(3) x N,''
JHEP {\bf 0301}, 077 (2003)
[arXiv:hep-th/0301110].
%%CITATION = HEP-TH 0301110;%%
}
\lref\tseyt{
A.~A.~Tseytlin,
``Conformal sigma models corresponding to 
gauged Wess-Zumino-Witten theories,''
Nucl.\ Phys.\ B {\bf 411}, 509 (1994)
[arXiv:hep-th/9302083].
%%CITATION = HEP-TH 9302083;%%
}

In this work we study some aspects
of the type 0 fermionic string~\foot{See \pol\ and references therein
for some background.} 
in a two dimensional black hole geometry~\foot{See \gpr\ and references
therein for a review.}.
The two dimensional black hole background corresponds to an exact
(to all orders in the inverse string tension $\alpha'$)
superconformal field theory (SCFT) on the worldsheet: the $SL(2)\over U(1)$
quotient SCFT. This is the supersymmetric worldsheet version of the bosonic 
$SL(2)\over U(1)$ black hole \refs{\witten,\dvv}.
In the Euclidean case, the axially gauged quotient sigma model 
is a semi-infinite ``cigar''
\efr\ with a non-trivial dilaton. 
Far from the tip of the cigar, the worldsheet theory is a sigma model on the 
infinite cylinder with a linear dilaton in the non-compact direction,
and two free fermions.
The string theory is weakly coupled in this asymptotic regime.
On the other hand,
the maximal value of the dilaton
is at the tip of the cigar; it is a finite, free parameter. 
Hence, this theory 
can be studied in a small string coupling ($g_s$) perturbation theory.  

The two dimensional Lorentzian background can be obtained either by 
gauging a different $U(1)$ in $SL(2)$, or by an analytic continuation
of the Euclidean energy on the cigar. This Wick rotation maps the 
cigar to the regime outside the horizon of the $1+1$ dimensional black hole.
The regime behind the singularity can be obtained by analytic continuation
of an axial/vector T-dual version of the cigar -- 
the ``trumpet'' \refs{\giv,\kir,\dvv} --
or by considering the analytic continuation of correlators of winding modes 
on the cigar to Lorentzian space-time.

Like the 2-d bosonic string, the closed string physical spectrum 
of the 2-d type 0 strings consists of 
massless ``tachyons''  (there are no string excitations in a two dimensional
target space-time~\foot{There are also discrete physical states 
appearing at special values of momenta, 
which we shall not consider in this work.}).
Hence, the type 0 2-d black hole is perturbatively stable.
Moreover, unlike the 2-d bosonic black hole but as in \refs{\tt,\six},
the type 0 black hole is also expected to be non-perturbatively well defined.
This fact on its own is already a good motivation to reconsider strings 
propagation in a 2-d black hole background, this time fermionic strings.

Another related motivation for this work is that similar to \refs{\tt,\six},
it is plausible that the type 0 2-d black hole has an open string dual:
a matrix quantum mechanics on its (unstable) localized D-branes. We shall
speculate on a possible direction to search for such a matrix model 
dual later.

The purpose of this work is to begin to collect some properties of
the 2-d type 0 string theories on $SL(2)\over U(1)$, 
and to consider its black hole interpretation.
In section 2,
we present some properties of the $SL(2)\over U(1)$ SCFT
which will be useful later. 
In section 3, we present physical operators in the type
0A and type 0B string theories. There are massless scalars (``tachyons'')
both in the NS-NS and R-R sectors of these theories.
In section 4,
we compute the two point functions of these scalars.
Then, in section 5,
we interpret these results as the reflection coefficients of scattering waves, 
which are incoming from an asymptotically flat regime of the black hole,
either outside the horizon or beyond its singularity,
and scattered either from the horizon or the singularity, respectively. 

In section 6,
we present our conjecture regarding a matrix quantum mechanics dual
to the type 0A string theory on the 2-d black hole. 
Finally, in the appendices we describe some of the technical details. 

\newsec{The $SL(2)/U(1)$ SCFT}

In this section we collect some properties of the two dimensional
superconformal field theory (SCFT) on ${SL(2)\over U(1)}$ which will
be useful later (see \pol\ for some preliminary background and
references therein, as well as \gk\ and references therein).

Consider first the superconformal WZW model on $SL(2)_{\k}$ 
at level $\k$. This theory has affine left-moving $SL(2)$ supercurrents,
\eqn\jjj{\psi^a+\theta\sqrt{2\over \k}J^a~,\qquad a=1,2,3,}
where 
\eqn\jjpp{J^a=j^a-{i\over 2}\epsilon^a_{~bc}\psi^b\psi^c~,}
and
\eqn\norms{\eqalign{\psi^a(z)\psi^b(z')
&\sim\frac{\eta^{ab}}{z-z'}~,
\qquad \eta^{ab}={\rm diag}(+,+,-)~, \cr &\cr
J^a(z)J^b(z')
&\sim\frac{\frac{\k}2\eta^{ab}}{(z-z')^2}+
\frac{i\epsilon^{ab}{}_cJ^c}{z-z'}~.}}
The purely bosonic currents $j^a$ generate an affine $SL(2)$
algebra at level $\k_B=\k+2$ and commute with the free fermions $\psi^a$, 
whereas the total (physical) currents $J^a$ generate a level $\k$ $SL(2)$
algebra and act on $\psi^a$ as follows from \jjpp,\norms.
Similarly, 
the theory has affine right-moving $SL(2)$ supercurrents, 
$\bar\psi^a+\bar\theta\sqrt{2\over\k}\bar J^a$.
This theory has a Virasoro central charge
\eqn\ccc{c(SL(2))={9\over 2}+{6\over \k}~.}
Define
\eqn\ppm{\psi^{\pm}={1\over\sqrt{2}}(\psi^1\pm i\psi^2)~,}
and introduce a canonically normalized scalar $H$, 
$H(z)H(z')\sim -\log(z-z')$, which is used to bosonize $\psi^{1,2}$:
\eqn\bos{\partial H\equiv \psi^2\psi^1=i\psi^-\psi^+~.}
Note that 
\eqn\jjh{J^3=j^3+i\partial H~,}
and
\eqn\note{j^3(z)\partial H(z')\sim 0~,\qquad 
J^3(z)i\partial H(z')\sim {1\over (z-z')^2}~.}
We introduce two other canonically normalized, independent scalars $X_3$ and
$X_R$, defined by
\eqn\jx{J^3=-\sqrt{\k\over 2}\partial X_3~,}
and
\eqn\ih{iH=\sqrt{2\over\k}X_3+i\sqrt{c\over 3}X_R~,}
where
\eqn\ccoset{c=c(SL(2)/U(1))=3+{6\over \k}~.}
The $X_3$ dependence in eq. \ih\ follows from \note\ and \jx.
Equations \jjh, \jx\ and \ih\ imply that
\eqn\jxx{j^3=-\sqrt{\k+2\over 2}\partial x_3~,}
where the canonically normalized scalar $x_3$ is:
\eqn\xthree{x_3=\sqrt{2\over\k}\left(\sqrt{\k+2\over 2}X_3+iX_R\right)~.}
Let $\Phi_{jm\bar m}$ be a primary field in the bosonic WZW model on 
$SL(2)_{\k_B}$ at level 
\eqn\kbk{\k_B=\k+2~,} 
which obeys
\eqn\jphi{j^3(z)\Phi_{jm\bar m}(z')\sim {m\Phi_{jm\bar m}(z')\over z-z'}~,}
and similarly $\bar m$ is the eigenvalue of the
right-moving current $\bar j^3$.
Since $\Phi_{jm\bar m}$ is purely bosonic -- independent off the free 
fermions $\psi^a$ -- it is also a primary of the superconformal theory on
$SL(2)_{\k}$ at level $\k$ with a $J^3$ eigenvalue equals to $m$:
\eqn\sjphi{J^3(z)\Phi_{jm\bar m}(z')\sim {m\Phi_{jm\bar m}(z')\over z-z'}~,}
and similarly for the right-movers.

We denote by $\Phi_{jm}(z)$ the holomorphic part of $\Phi_{jm\bar m}$.
Its scaling dimension is:
\eqn\scdim{h(\Phi_{jm})=-{j(j+1)\over \k}~.}
Using eqs. \jx\ and \jxx,\xthree, we can decompose:
\eqn\dec{\Phi_{jm}=U_{jm}e^{m\sqrt{2\over\k}X_3}
=V_{jm}e^{m\sqrt{2\over\k+2}x_3}
=V_{jm}e^{i{2m\over \k+2}\sqrt{c\over 3}X_R}e^{m\sqrt{2\over\k}X_3}~,}
where $V_{jm}$ is a primary of the bosonic Euclidean quotient CFT on 
$SL(2)_{\k_B}\over U(1)$ and
$U_{jm}$ is a primary of the superconformal quotient 
$SL(2)_{\k}\over U(1)$. The Virasoro 
central charge $c$ of the latter is given in eq. \ccoset.
The scaling dimensions of $V$ and $U$ are
\eqn\hv{h(V_{jm})=-{j(j+1)\over \k}+{m^2\over \k+2}~,}
and
\eqn\hu{h(U_{jm})={-j(j+1)+m^2\over \k}~.}
Similarly, using \ih, we can decompose
\eqn\expnh{e^{inH}=e^{n(\sqrt{2\over\k}X_3+i\sqrt{c\over 3}X_R)}~.}
Operators in the Euclidean ${SL(2)_{\k}\over U(1)}$ 
SCFT are obtained from their 
``parent'' operators in the $SL(2)_\k$ SCFT by eliminating the $X_3$ and 
$\psi^3$ dependence.
Hence, primary fields in the ${SL(2)_{\k}\over U(1)}$ SCFT take the form:
\eqn\vjmn{V_{jm}^n=V_{jm}e^{i({2m\over \k+2}+n)\sqrt{c\over 3}X_R}~.}
Their scaling dimension is:
\eqn\hvjmn{h(V_{jm}^n)=-{j(j+1)\over \k}+{m^2\over \k+2}
+{c\over 6}\left(n+{2m\over \k+2}\right)^2=
{-j(j+1)+(m+n)^2\over\k}+{n^2\over 2}~.} 
In the Neveu-Schwarz (NS) sector $n$ is an integer while in the Ramond (R) 
sector $n$ is in $Z+{1\over 2}$:~\foot{This follows, for instance,
from the fact that 
the NS (R) sector in a quotient SCFT is induced from the NS (R) sector
of the ``parent'' theory, and from comment (4) below.
It also follows from comment (6) bellow.}
\eqn\nnsr{\eqalign{
NS:& \qquad n\in Z~,\cr
R: & \qquad n\in Z+{1\over 2}~.}}
A few comments are in order:
\item{(1)}
The SCFT on ${SL(2)_{\k}\over U(1)}$ has an $N=(2,2)$ superconformal symmetry.
The holomorhic $U(1)$ $R$-current of the left-moving $N=2$ algebra is
\eqn\jr{J_R=i\sqrt{c\over 3}\partial X_R~,}
and similarly for the right-moving algebra.
\item{(2)}
Equations \jr\ and \vjmn\ imply that the $R$-charge of $V_{jm}^n$ is:
\eqn\qr{Q_R(V_{jm}^n)={c\over 3}
\left({2m\over \k+2}+n\right)={2m\over \k}+{nc\over 3}~.}
\item{(3)}
When $n=0$:
\eqn\neqo{V_{jm}^0=U_{jm}=V_{jm}e^{i({2m\over \k+2})\sqrt{c\over 3}X_R}~,}
where $V_{jm}$ and $U_{jm}$ are given above (see eqs. \dec,\hv,\hu,\vjmn). 
\item{(4)}
The ``parent'' operator of $V_{jm}^n$ is the $SL(2)_\k$ 
operator~\foot{In the R sector ($n\in Z+{1\over 2}$), we have ignored
the spin field $s^3$ of the $\psi^3$ fermion in $SL(2)$, 
which is eliminated anyway in the quotient SCFT.}
\eqn\parent{\Phi_{jm}e^{inH}=V_{jm}^ne^{\sqrt{2\over\k}(m+n)X_3}~.}
Hence, its $J^3$ eigenvalue is $m+n$.
\item{(5)}
The Lorentzian ${SL(2)_\k\over U(1)}$ quotient is obtained by gauging a
space-like direction instead of the time-like direction generated by $J^3$.
Hence, comment (4) implies that the analytic continuation from the Euclidean 
black hole to the Lorentzian 
one is done by a Wick rotation of the real $m+n$ 
eigenvalue to an imaginary value:
\eqn\mie{\eqalign{
Euclidean:&\qquad m+n={K\over 2}~, \qquad K\in R~,\cr
Lorentzian:&\qquad m+n={iP\over 2}~, \qquad P\in R~.
}}
\item{(6)}
The operators obtained by setting $n=\pm {1\over 2}$ and
taking $V_{jm}\rightarrow 1$ (which has $j=m=0$) on the right hand side 
of eq. \vjmn\ are the spin fields:
\eqn\spm{{\cal S}^{\pm}=e^{\pm {1\over 2}\int J_R}
=e^{\pm {i\over 2}\sqrt{c\over 3}X_R}~.}
These generate the Ramond ground states when acting on the NS vacuum.
\item{(7)}
The geometry of the Euclidean $SL(2)_k\over U(1)$ SCFT background is the 
following \refs{\efr,\witten}. 
The metric and dilaton are~\foot{Here we work with $\alpha'=2$.}
\eqn\metric{ds^2=\k(dr^2+\tanh^2rd\theta^2)~,\qquad r\geq 0~,\qquad
\theta\simeq \theta+2\pi~,}
and 
\eqn\dilaton{e^{\bf\Phi}={e^{{\bf\Phi}_0}\over\cosh r}~,}
where the dilaton ${\bf\Phi}$ is normalized such that the string coupling is 
$g_s=e^{\langle{\bf\Phi}\rangle}$, and ${\bf\Phi}_0$ is a constant.
This is the cigar shaped background, which is exact to all orders in 
$\alpha'$ (up to an overall constant normalization) \tseyt.
\item{(8)}
Asymptotically away from the tip of the cigar this background becomes 
the cylinder $R_\phi\times S^1_x$ 
(parametrized by two canonically normalized scalars $x$ and $\phi$)
with two free fermions $\psi_\phi$, $\psi_x$ and a linear dilaton:
\eqn\metas{ds^2={1\over 2}(d\phi^2+dx^2)~, \qquad 
{\bf\Phi} = -{Q\over 2}\phi~,}
where~\foot{In the particular case $Q=2$ ($\k={1\over 2}$) we have:
$\phi=r$ and $x=\theta$.
Incidentally, we shall restrict to this particular case in the next sections.}
\eqn\rphiq{Q=\sqrt{2\over \k}~, \qquad \phi={2\over Q}r~, \qquad 
x={2\over Q}\theta~, \qquad x\simeq x+{4\pi\over Q}~.}
In the asymptotic cylinder the central charge reads:
\eqn\cas{c=3(1+Q^2)~,}
and the $U(1)_R$ current of the $N=2$ algebra is:
\eqn\ras{J_R=i\partial(H'+Qx)~,}
where the canonically normalized scalar $H'$ is used to bosonize the 
free fermions:
\eqn\hprbos{\partial H'=\psi_x\psi_\phi~.} 
\item{(9)}
The asymptotic behavior of the SCFT primaries $U_{jm}$ is:
\eqn\uas{U_{jm}\rightarrow e^{iQmx}e^{Qj\phi}~.}
Its $R$-charge is thus
$Q_R=Q^2m={2m\over \k}$ (in agreement with eqs. \qr,\neqo).
Moreover, 
eqs. \jr\ and \ras\ imply that the asymptotic behavior of the scalar
$X_R$ is:
\eqn\xras{\sqrt{c\over 3}X_R\rightarrow H'+Qx~,}
and thus, using eqs. \vjmn,\neqo,\uas\ and \xras, one finds that the 
asymptotic behavior of the other $SL(2)\over U(1)$ superconformal primaries
is~\foot{More precisely, $V_{j,m,\bar m}^{n,\bar n}\to
e^{i(nH'-\bar n\bar H')}e^{iQ[(m+n)x-(\bar m +\bar n)\bar x]}
\left(e^{Qj\phi}+R(j,m,\bar m)e^{-Q(j+1)\phi}\right)$ 
(see \grs\ and references therein), and it is the ``reflection coefficient''
$R(j,m,\bar m)$ that we shall actually compute later.}:
\eqn\vjmnas{V_{jm}^n\rightarrow e^{inH'}e^{iQ(m+n)x}e^{Qj\phi}.}  
Its scaling dimension is 
$h(j,m,n)={Q^2\over 2}\left(-j(j+1)+(m+n)^2\right)+{n^2\over 2}$
(in agreement with eq. \hvjmn); here we used:
\eqn\heap{h(e^{\alpha\phi})=-{1\over 2}\alpha(\alpha+Q)~.}
\item{(10)}
In the Euclidean quotient:
\eqn\mmbar{m+n={1\over 2}(p+w\k)~, \qquad 
\bar m +\bar n =-{1\over 2}(p-w\k)~, \qquad
p,w\in Z~,}
where $p$ and $w$ are momenta and winding numbers on the 
cigar geometry~\foot{The reason that we have
$(m+n,\bar m +\bar n)={1\over 2}(p+w\k,-p+w\k)$ 
is due to the fact that the cigar is obtained
by an {\it axial} $U(1)$ gauging of $SL(2)$ \dvv; $(m+n,\bar m +\bar n)$ 
are the $(J^3,\bar J^3)$ eigenvalues.
A vector gauging leads to the T-dual trumpet \refs{\giv,\kir,\dvv}, 
where $\bar m +\bar n\rightarrow -(\bar m +\bar n)$.
As a consequence, our conventions regarding type 0A versus 0B
for the cigar are T-dual to what is used in Liouville.}.
This is manifest, for instance, from the asymptotic behavior \vjmnas:
eqs. \vjmnas\ and \mmbar\ imply that the left and right-moving
momenta on the cigar are
$$(k,\bar k)=Q(m+n,\bar m+\bar n)={1\over\sqrt{2}}\left(
{p\over\sqrt\k}+w\sqrt\k,-{p\over\sqrt\k}+w\sqrt\k\right)~,$$
which is compatible with the fact that the asymptotic radius of the cigar is
$${R\over\sqrt{\alpha'}}=\sqrt\k~.$$
\item{(11)}
Unitarity implies \refs{\gk,\pak} that either 
\eqn\jreal{-{1\over 2}<j<{k-1\over 2}~, \qquad j\in R~,}
corresponding to states (in the discrete series)
localized near the tip of the cigar,
or
\eqn\jim{j\in -{1\over 2}+is~, \qquad s\in R~,}
corresponding to scattering states (in the continuous series).

\newsec{Physical Operators in Type 0 String Theory on $SL(2)\over U(1)$} 

The fermionic string~\foot{For some background needed for this section 
see \pol\ and references therein.}
has a critical central charge $c=15$.
Hence, consistency of a fermionic string propagating in an 
$SL(2)\over U(1)$ SCFT background requires that (see eq. \ccoset):
\eqn\khalf{\k={1\over 2}\quad \Leftrightarrow\quad  c(SL(2)_{1/2}/U(1))=15~.}
We thus consider the fermionic string on the $SL(2)_{1/2}\over U(1)$ 
superconformal worldsheet theory.  

Using the results of the previous section, and standard arguments applied to
a fermionic string theory in a two dimensional space-time
\refs{\dfk,\pol}, one finds that
the physical operators have the following form 
(we first consider the holomorhic part of
the operators~\foot{The way the left-moving part is combined with the 
right-moving part depends on which type of fermionic string is 
being considered; this will be discussed below.}).
In the NS sector:
\eqn\tns{T^\pm(k)=
e^{-\varphi}U_{-{1\over 2}\pm {k\over 2},{k\over 2}}=e^{-\varphi}
V_{-{1\over 2}\pm {k\over 2},{k\over 2}}e^{i{2k\over \sqrt{5}}X_R}~,}
where $T^+(k)$ obey the unitarity bound \jreal\ when $k>0$
and $T^-(k)$ obey the bound when $k<0$. 
In the Euclidean case (see eq. \mmbar\ with $\k={1\over 2}$, $m={k\over 2}$ 
and $n=0$):
\eqn\keuc{k=p+{w\over 2}~, \qquad p,w\in Z~,}
while in the Lorentzian case:
\eqn\klor{k=iP~,\qquad  P\in R~.}
We chose to write the NS operators \tns\ in the $-1$ picture;
$\varphi$ is the scalar field arising in the bosonized
superghost system of the worldsheet supersymmetry.
In the R sector:
\eqn\sr{S^{\pm}(k)=e^{-{\varphi\over 2}}
V_{-{1\over 2}\pm{k\over 2},{k\over 2}\mp{1\over 2}}^{\pm{1\over 2}}=
e^{-{\varphi\over 2}}V_{-{1\over 2}\pm{k\over 2},{k\over 2}\mp{1\over 2}}
e^{{i\over\sqrt{5}}(2k\pm{1\over 2})X_R}~.}
Again, $S^{+}(k)$ obey the unitarity bound \jreal\ when $k>0$
and $S^{-}(k)$ obey the bound when $k<0$, $k$ is given in \keuc\ or \klor\
(see eq. \mmbar\ with $\k={1\over 2}$, $m={k\over 2}\mp{1\over 2}$ 
and $n=\pm{1\over 2}$),
and we chose to write the R sector operators in the $-{1\over 2}$ picture.

A few comments are in order:
\item{(i)}
One can verify that
\eqn\verify{h(T^{\pm}(k))=h(S^{\pm}(k))=1~,}
by using \hu,\hvjmn\ and
\eqn\heqvp{h(e^{q\varphi})=-{1\over 2}q(q+2)~.}
\item{(ii)}
The asymptotic behavior of these vertex operators 
at the weak coupling regime is (see also \murthy):
\eqn\tas{T^{\pm}(k)\rightarrow e^{-\varphi}e^{ikx}e^{(-1\pm k)\phi}~,}
where $\phi$ is a canonically normalized scalar with a background
charge $Q=2$ (${\bf\Phi}=-\phi$) and
$x$ is a canonically normalized scalar with periodicity:
\eqn\xxpi{x\simeq x+2\pi~,}
and
\eqn\sas{S^{\pm}(k)\rightarrow e^{-{\varphi\over 2}}
e^{\pm {i\over 2}H'}e^{ikx}e^{(-1\pm k)\phi}~.}
Here we used $Q=2$ and $c=15$ in eqs. \rphiq,\uas,\xras\ and \vjmnas.
\item{(iii)}
The operators \tas\ and \sas\ are BRST invariant.
This is shown in appendix A.
\item{(iv)}
The $N=2$ $R$-current is 
\eqn\rcur{J_R=i\sqrt{5}\partial X_R\rightarrow i\partial (H'+2x)~,}
and the spin field is:
\eqn\spf{{\cal S}^{\pm}
=e^{\pm {1\over 2}\int J_R}=e^{\pm {i\over 2}\sqrt{5}X_R}
\rightarrow e^{\pm {i\over 2}(H'+2x)}~.}
The right hand side in eqs. \rcur\ and \spf\ is the asymptotic behavior.
\item{(v)}
The analysis for the anti-holomorhic part of the vertex operators is obvious, 
and follows the same lines as above, but now for the right-movers.
In Euclidean space-time the left and right-moving ``momenta'' 
on the cigar are
(see eq. \mmbar\ and use the same arguments which led to \keuc):  
\eqn\krm{(k,\bar k)=\left(p+{w\over 2},-p+{w\over 2}\right)~, 
\qquad p,w\in Z~.} 
\item{(vi)}
Plugging the values $j=-{1\over 2}+i{P\over2}$ (corresponding to 
scattering states \jim) and $m+n=i{E\over 2}$ 
(corresponding to a Wick rotation to the Lorentzian black hole \mie) 
into $U_{jm}$ and $V_{jm}^{n}$, we see that the on-shell conditions 
\verify\ translate into the condition:
\eqn\massless{E^2-P^2=0~.} 
Hence, after combining left-movers with right-movers, the physical 
vertex operators correspond to massless fields in the $1+1$ dimensional 
space-time.

\vskip .1in
\noindent
We are now prepared to construct the type 0 theories.
This requires to combine left-movers with right-movers in a way that
is tree-level unitary and one-loop modular invariant.
There are four non-equivalent ways to do it:

\item{(1)}
{\bf Type 0B}: 

\noindent
We include in the theory the two (mutually local) operators
$Q^{++}$ and $Q^{--}$, 
\eqn\sspp{Q^{\pm\pm}(z,\bar z)
=e^{-{\varphi\over 2}-{\bar\varphi\over 2}}\SS^{\pm}\bar\SS^{\pm}~,}
where the spin fields $\SS^{\pm}$ are given in \spf.
Mutual locality with these operators is the analog of the diagonal GSO
projection in flat space.
This mutual locality implies that~\foot{More generally,
mutual locality implies that $n-\bar n\in 2Z$. However, since the $2$-d string
theory has no physical excitations, the operators $T(k)$ and $S(k)$ only
have $n=0$ and $n=\pm{1\over 2}$, respectively.} 
\eqn\nbarn{n=\bar n~.}
Hence, there are physical operators only in the NS-NS and R-R sectors.
The on-shell condition further implies (see \hvjmn) that 
\eqn\mnmn{(m+n)^2=(\bar m+\bar n)^2~.}
Hence, in the NS-NS sector ($n=\bar n=0$) we have:
\eqn\mmnsns{NS-NS:\qquad m=\pm\bar m~.}
On the other hand, in the R-R sector ($n=\bar n=\pm {1\over 2}$), eq. \mnmn\
now implies:
\eqn\mmrr{R-R:\qquad m=\bar m~.}
All in all, we learn that the physical operators in the NS-NS sector are:
\eqn\nsphys{{\bf T}^{\pm\e}(k)=
e^{-\varphi-\bar\varphi}
V^{0,0}_{-{1\over 2}\pm {k\over 2},{k\over 2},\e{k\over 2}}
=e^{-\varphi-\bar\varphi}
V_{-{1\over 2}\pm {k\over 2},{k\over 2},\e{k\over 2}}
e^{i{2k\over \sqrt{5}}(X_R(z)-\e \bar X_R(\bar z))}~, \qquad \e=\pm~,}
where, in the Euclidean case,  
${\bf T}^{+\e}(k)$ obey the unitarity bound \jreal\ when $k>0$
and ${\bf T}^{-\e}(k)$ obey the bound when $k<0$, and
\eqn\keke{\eqalign{
k=\bar k={w\over 2}\in {Z\over 2}&\qquad if \qquad \e=1~,\cr
k=-\bar k=p\in Z &\qquad if \qquad \e=-1~.}}
The physical operators in the R-R sector are:
\eqn\srr{\eqalign{
{\bf S}^{\pm}(k)=&e^{-{\varphi\over 2}-{\bar\varphi\over 2}}
V_{-{1\over 2}\pm{k\over 2},{k\over 2}\mp{1\over 2},
{k\over 2}\mp{1\over 2}}^{\pm{1\over 2},\pm{1\over 2}}\cr 
=&e^{-{\varphi\over 2}-{\bar\varphi\over 2}}
V_{-{1\over 2}\pm{k\over 2},{k\over 2}\mp{1\over 2},
{k\over 2}\mp{1\over 2}}
e^{{i\over\sqrt{5}}(2k\pm{1\over 2})(X_R(z)-\bar X_R(\bar z))}~,}}
where, in the Euclidean case,  
${\bf S}^+(k)$ obey the unitarity bound \jreal\ when $k>0$
and ${\bf S}^-(k)$ obey the bound when $k<0$, and
\eqn\kkw{k=\bar k={w\over 2}~,\qquad w\in Z~.}
The operators $V_{jm\bar m}^{n\bar n}$ in \nsphys\ and \srr\ are the
combinations of the left-moving primaries $V_{jm}^n$ in \vjmn\ with
the right-moving $\bar V_{j\bar m}^{\bar n}$.
In both the NS-NS and the R-R sectors, we have:
\eqn\mnk{m+n={k\over 2}~.}
On the other hand, for the right-moving $\bar J^3$ number we have:
\eqn\mnbark{\eqalign{
\bar m+\bar n=&\pm {k\over 2}\qquad in\quad NS-NS~,\cr
\bar m+\bar n=&{k\over 2}\qquad in\quad R-R~.}}
Hence, the analytic continuation from the Euclidean cigar to the Lorentzian
black hole is done by Wick rotating (see \mie):
\eqn\ktop{k\rightarrow iP~,\qquad P\in R~.}

\item{(2)}
{\bf Type 0A}: 

\noindent
We include in the theory the two (mutually local) operators
$Q^{+-}$ and $Q^{-+}$, 
\eqn\sspm{Q^{\pm\mp}(z,\bar z)
=e^{-{\varphi\over 2}-{\bar\varphi\over 2}}\SS^{\pm}\bar\SS^{\mp}~,}
where the spin fields $\SS^{\pm}$ are given in \spf.
Mutual locality now implies that~\foot{More generally,
mutual locality implies that $n+\bar n\in 2Z$, but $T(k)$ and $S(k)$ only
have $n=0$ and $n=\pm{1\over 2}$, respectively.}
\eqn\nmbarn{n=-\bar n~,}
and again, there are physical operators only in the NS-NS and R-R sectors.
As before, the on-shell condition further implies eq. \mnmn. 
Hence, in the NS-NS sector ($n=\bar n=0$) we have again the relation \mmnsns,
but in the R-R sector ($n=-\bar n=\pm {1\over 2}$), eq. \mnmn\
now implies:
\eqn\mmmrr{R-R:\qquad m=-\bar m~.}
Thus, the physical operators in the NS-NS sector are the same as in type 0B
\nsphys,\keke, while in the R-R sector the physical operators are:
\eqn\srra{\eqalign{
{\bf \tilde S}^{\pm}(k)=& e^{-{\varphi\over 2}-{\bar\varphi\over 2}}
V_{-{1\over 2}\pm{k\over 2},{k\over 2}\mp{1\over 2},
-{k\over 2}\pm{1\over 2}}^{\pm{1\over 2},\mp{1\over 2}}\cr
=& e^{-{\varphi\over 2}-{\bar\varphi\over 2}}
V_{-{1\over 2}\pm{k\over 2},{k\over 2}\mp{1\over 2},
-{k\over 2}\pm{1\over 2}}
e^{{i\over\sqrt{5}}(2k\pm{1\over 2})(X_R(z)+\bar X_R(\bar z))}~,
}}
where, in the Euclidean case, 
\eqn\kkp{k=-\bar k=p~,\qquad p\in Z~.}
Again, both in the NS-NS and R-R sectors eq. \mnk\ is satisfied. 
On the other hand, for the right-moving numbers we now have:
\eqn\mnbark{\eqalign{
\bar m+\bar n=&\pm {k\over 2}\qquad in\quad NS-NS~,\cr
\bar m+\bar n=&-{k\over 2}\qquad in\quad R-R~.
}}
Therefore, the analytic continuation to the Lorentzian
black hole is done again by Wick rotating as in \ktop.

\item{(3)}
{\bf Type IIB}: 

\noindent
Here we require mutual locality with the holomorhic operators $Q^+(z)$
and $\bar Q^+(\bar z)$ (or $Q^-(z)$ and  $\bar Q^-(\bar z)$), where
\eqn\qpqpqmqm{Q^{\pm}(z)=e^{-{\varphi\over 2}}\SS^{\pm}(z)~,}
and similarly for the right movers \refs{\daps,\kuse,\murthy}.
This is the analog of a chiral GSO projection in flat space, and it leads 
to the supersymmetric 2-d black hole (in the Euclidean case).
We shall not consider this theory here,
%~\foot{Due to the chiral nature of the
%projection, the geometry of space-time is not associated manifestly 
%with a two dimensional black hole. THIS COMMENT IS TRUE FOR SL(2)/U(1)
%x D-dim minkowski w/ D>0, BUT NOT HERE}, 
except for two remarks:
operators in the NS-NS and R-R sectors correspond to space-time bosons,
while operators in the NS-R and R-NS sectors correspond to space-time 
fermions.
However, there is no boson-fermion degeneracy in the spectrum, because there
is no Poincar\'e invariance in space-time.
Some evidence in favor of a supersymmetric matrix model dual 
to the superstrings in the 2-d black hole geometry was presented recently
in \mgmv.  

\item{(4)}
{\bf Type IIA}: 

\noindent
Here we impose an opposite chiral projection on the left and right-movers, 
namely, we include the holomorhic operators $Q^+(z)$
and $\bar Q^-(\bar z)$ (or $Q^-(z)$ and  $\bar Q^+(\bar z)$), where
$Q^{\pm}(z)$ are given in \qpqpqmqm. We shall not consider this theory here,
except for the remarks in the type IIB case which are correct in the IIA case
as well. 

\newsec{The Two Point Functions}

The physical operators ${\bf T}(k)$, ${\bf S}(k)$ and ${\bf \tilde S}(k)$
are of the form:
\eqn\vform{V_{jm\bar m;q\bar q}(k_R,\bar k_R)
=e^{q\varphi+\bar q\bar\varphi}
V_{jm\bar m}e^{i(k_RX_R-\bar k_R\bar X_R)}~.}
Hence, the two point functions (2-p-f) take the form~\foot{We 
suppress the $z,\bar z$ dependence.}:
\eqn\tpf{\eqalign{
&\langle V_{j'm'\bar m';q'\bar q'}(k'_R,\bar k'_R)
V_{jm\bar m;q\bar q}(k_R,\bar k_R) \rangle\cr
&=\langle e^{q'\varphi+\bar q'\bar\varphi}e^{q\varphi+\bar q\bar\varphi}
\rangle
\langle e^{i(k'_RX_R-\bar k'_R\bar X_R)}e^{i(k_RX_R-\bar k_R\bar X_R)}\rangle
\langle V_{j'm'\bar m'}V_{jm\bar m}\rangle~.
}}
Unlike flat target-space, in $SL(2)\over U(1)$
the 2-p-f on the sphere do not necessarily vanish 
\refs{\gk,\gktwo}. Conservation laws require though that the 2-p-f
vanish unless:
\eqn\conserv{\eqalign{
&k_R+k'_R=0=\bar k_R+\bar k'_R~,\cr
&q+q'=-2=\bar q+\bar q'~,\cr
&m+m'=0=\bar m+\bar m'~, \qquad j=j'~.
}}
The first two correlators on the r.h.s of eq. \tpf\ are trivial:
they are equal to $1$ when \conserv\ is satisfied.  
The third two point function is a 2-p-f in bosonic $SL(2)\over U(1)$ at level 
$\k_B=\k+2$. When \conserv\ is satisfied, it is equal to 
\refs{\gk,\gktwo}:~\foot{Note that (in the Euclidean cigar) 
$\langle V_{j,-m,-\bar m}V_{jm\bar m}\rangle=
\langle V_{jm\bar m}V_{j,-m,-\bar m}\rangle$, as it should,
since $m-\bar m=p\in Z$;
this can be verified by using properties of $\GG$ functions
on the right hand side.
On the other hand, 
$\langle V_{-(j+1),-m,-\bar m}V_{-(j+1),m,\bar m}\rangle=
\langle V_{j,-m,-\bar m}V_{jm\bar m}\rangle^{-1}$.}
\eqn\sltpf{\langle V_{j,-m,-\bar m}V_{jm\bar m}\rangle= 
\nu(\k)^{2j+1}{\Gamma(1-{2j+1\over\k})\over \Gamma(1+{2j+1\over\k})}
{\Gamma(-2j-1)\over \Gamma(2j+1)}
{\Gamma(j+m+1)\Gamma(j-\bar m+1)\over\Gamma(-j-\bar m)\Gamma(-j+m)}~.}
The $j,m,\bar m$-independent constant $\nu$ is the analog of the cosmological
constant $\mu$ in Liouville theory (more precisely, a sine-Liouville; 
see eq. (4.89) in \gkthree).
{}From now on we choose~\foot{See \gkthree\ for a discussion on the
freedom to make such a choice.}:
\eqn\nuone{\nu(\k)=1~.}
We shall now use these results to compute the 2-p-f of NS-NS and R-R
scalars in the type~0 string theories on the $2$-d black hole background.

\subsec{The 2-P-F of NS-NS Scalars in Type 0B and 0A}

The physical spectrum of NS-NS scalars is identical in the type 0B 
and type 0A theories. In the Euclidean cigar, it consists of states with 
$m=-\bar m={k\over 2}={p\over 2}$ 
($\e=-$ in \nsphys) and $m=\bar m={k\over 2}={w\over 4}$ $(\e=+$ in \nsphys).
Recall that the integers $p$ and $w$ are momentum and winding numbers on the 
cigar background.

For $m=\bar m$ it is convenient to define:
\eqn\ttkk{\eqalign{
T(k)=\T^{++}(k) \qquad k>0~,\cr
T(k)=\T^{-+}(k) \qquad k<0~,
}}
and for $m=-\bar m$:
\eqn\ttttkk{\eqalign{
\tilde T(k)=\T^{+-}(k) \qquad k>0~,\cr
\tilde T(k)=\T^{--}(k) \qquad k<0~.
}}
Following the details in appendix B, the two point functions are:
\eqn\tktk{\eqalign{
k={w\over 2}\in {Z\over 2}~, \qquad m &=\bar m:\cr
\langle T(-k)T(k)\rangle &={\GG(1-2|k|)\GG(-|k|)\GG({1\over 2}+|k|)\over
\GG(1+2|k|)\GG(|k|)\GG({1\over 2}-|k|)}~,
}}
and~\foot{Note that for $k=p\in Z$ we have $\cos(\pi p)=1$, 
hence $\langle \tilde T(-p)\tilde T(p)\rangle=
\langle\tilde T(p)\tilde T(-p)\rangle=\langle T(-p)T(p)\rangle$,
and there are double poles in \tktk\ and here.
We keep however the present form of 
$\langle \tilde T(-k)\tilde T(k)\rangle$
since we are interested in its analytic continuation.}
\eqn\ttkttk{\eqalign{
k=p\in Z~, \qquad m&=-\bar m:\cr
\langle \tilde T(-k)\tilde T(k)\rangle &=
\left({\GG(1-2|k|)\GG(-|k|)\over\GG(1+2|k|)\GG(|k|)}\right)
\left({\GG({1\over 2}+k)\over\GG({1\over 2})}\right)^{2sign(k)}\cr 
&=\langle T(-k)T(k)\rangle
\left({1\over \cos(\pi k)}\right)^{sign(k)}~.
}}
After analytic continuation $|k|\to iP$, $P\in R$, 
to the $1+1$ Lorentzian black hole, we have:
\eqn\rsp{\left(R_s(P)\right)^{sign(P)}\equiv
\langle T(-iP)T(iP)\rangle=e^{i\varphi(P)}~,\qquad P\in R~,
}
where the phase $\varphi(P)$ is
\eqn\phase{i\varphi(P)=\log\left({
\GG(1-2iP)\GG(-iP)\GG({1\over 2}+iP)\over
\GG(1+2iP)\GG(iP)\GG({1\over 2}-iP)}\right)~,
}
and
\eqn\rhp{\left(R_h(P)\right)^{sign(P)}\equiv
\langle \tilde T(-iP)\tilde T(iP)\rangle
=e^{i\varphi(P)}\left({1\over\cosh(\pi P)}\right)^{sign(P)}~,\qquad P\in R~.
}

\subsec{The 2-P-F of R-R Scalars in Type 0B}

%{\bf xxx REMARK: 
%
%Before imposing BRST invariance there are four operators with $h=\bar h1$
%in the $-{1\over 2}$ picture:
%\eqn\srree{\eqalign{
%{\bf S}^{\e_1\e_2}(k)=&e^{-{\varphi\over 2}-{\bar\varphi\over 2}}
%V_{-{1\over 2}+\e_1{k\over 2},{k\over 2}-\e_2{1\over 2},
%{k\over 2}-\e_2{1\over 2}}^{\e_2{1\over 2},\e_2{1\over 2}}\qquad \qquad
%\e_{1,2}=\pm\cr 
%=&e^{-{\varphi\over 2}-{\bar\varphi\over 2}}
%V_{-{1\over 2}+\e_1{k\over 2},{k\over 2}-\e_2{1\over 2},
%{k\over 2}-\e_2{1\over 2}}
%e^{{i\over\sqrt{5}}(2k+\e_2{1\over 2})(X_R(z)+\bar X_R(\bar z))}~.}}
%We found that BRST invariance implies $\e_1=\e_2$ leading to \sr,\srr\ 
%and what follows it (and similarly to \srra\ and what follows it).
%However, if we made a sign error, and actually BRST invariance implies
%the opposite condition, $\e_1=-\e_2$, then the bulk of this and the next
%subsections should be interchanged with the footnotes. 
%xxx}

The physical spectrum in the R-R sector of the type 0B string theory 
on the cigar consists of winding modes \kkw,\mnk: 
$m+n=\bar m +\bar n={k\over 2}={w\over 4}$.
It is convenient to define:
\eqn\spksmk{\eqalign{
S(k)=\S^{+}(k) \qquad k>0~,\cr
S(k)=\S^{-}(k) \qquad k<0~,
}}
where $S(k)$ is either in the $(-{1\over 2},-{1\over 2})$ or the 
$(-{3\over 2},-{3\over 2})$ picture (see appendices A and C).
Following the details in appendix C, the two point functions are:
\eqn\sstpf{\langle S(-k)S(k)\rangle=
{\GG(1-2|k|)\GG(-|k|)\over \GG(1+2|k|)\GG(1-|k|)\GG(0)}=
{2(-)^{w}\over (|w|!)^2}~,}
(or its inverse).
The analytic continuation to scattering states in
Lorentzian space gives (see appendix C):
\eqn\rzero{R_s(P)=0~, \qquad \forall~P~,}
(or its inverse).

\subsec{The 2-P-F of R-R Fields in Type 0A}

The physical spectrum in the R-R sector of the type 0A string theory 
on the cigar consists of momentum modes \kkp,\mnbark: 
$m+n=-\bar m -\bar n={k\over 2}={p\over 2}$. 
It is convenient to define:
\eqn\tspktsmk{\eqalign{
\tilde S(k)=\tilde\S^{+}(k) \qquad k>0~,\cr
\tilde S(k)=\tilde\S^{-}(k) \qquad k<0~,
}}
where $\tilde S(k)$ is either in the $(-{1\over 2},-{1\over 2})$ or the 
$(-{3\over 2},-{3\over 2})$ picture (see appendices A,C and D).
Following the details in appendix D, the two point functions are:
\eqn\ppp{\langle \tilde S(-k)\tilde S(k)\rangle=
{2(-)^{p-1}\over (|2p|!)^2}~,}
(or its inverse).
They are equal to the 2-p-f of momentum modes
in the NS-NS sector, up to ``leg factors:''
\eqn\tststpf{\langle \tilde S(-k)\tilde S(k)\rangle=
\langle\tilde T(-k)\tilde T(k)\rangle
\left({\GG({1\over 2}+k)\over\GG({1\over 2})}\right)^{2}
\left({\GG(1)\over\GG(1+k)}\right)^{2}~, 
\qquad k<0~,}
\eqn\tststpff{\langle \tilde S(-k)\tilde S(k)\rangle=
\langle\tilde T(-k)\tilde T(k)\rangle
\left({\GG({1\over 2})\over\GG({1\over 2}+k)}\right)^{2}
\left({\GG(k)\over\GG(0)}\right)^{2}~,\qquad k>0~,}
(or its inverse), where $\langle\tilde T(-k)\tilde T(k)\rangle$
is given in \ttkttk.
Recall that actually \tststpff\ $=$ \tststpf\ $=$ \ppp.

The formal analytic continuation of \tststpf\
to scattering states in Lorentzian space gives~\foot{For $P<0$; 
see appendix D
and recall that we choose the branch with $-i(j+{1\over 2})>0$.}:
\eqn\rzeroo{\left(R_h(P)\right)^{-1}\equiv
\langle \tilde S(-iP)\tilde S(iP)\rangle=
\langle\tilde T(-iP)\tilde T(iP)\rangle
\left({\GG({1\over 2}+iP)\over\GG({1\over 2})}\right)^{2}
\left({\GG(1)\over\GG(1+iP)}\right)^{2}~,}
or its inverse~\foot{For $P>0$.}, where
$\langle\tilde T(-iP)\tilde T(iP)\rangle$ is given in \rhp.
On the other hand, the analytic continuation of
\tststpff\ is $0$ for all $P$.

\newsec{Scattering from the Black Hole Horizon and Singularity}

In this section we present the interpretation of the results 
in the previous section in terms of the black hole physics.

\subsec{Scattering of NS-NS Scalars}

The two point functions are normalized such that,
after analytic continuation, they give the 
reflection coefficients~\foot{Or the inverse of $R(P)$, 
depending on whether the scattering wave is incoming or outgoing, namely,
on the relative sign of energy and momentum in the scattering state.
This relative sign is changed when we take 
$(j,m,\bar m)\to (-(j+1),m,\bar m)$: it takes 
$P=-2i(j+{1\over 2})\to -P$ and does not change $E=-i(m\pm\bar m)$
($\pm$ sign if $m=\pm\bar m$; see \egkr,\grs).} 
$R(P)$ of scattering waves, which are incoming 
from an asymptotically flat regime in the extended eternal 
black hole~\foot{This can be shown,
for instance, by following sections 3 and 4 of \grs,
and references therein.}. 

The scattering of momentum modes in the cigar background has the following
interpretation. It is analytically 
continued to the scattering of momentum modes incoming from the 
asymptotically flat regime outside the black hole horizon, 
and scattered from the horizon.
The result \rhp\ shows that the reflection coefficient $R_h(P)$ satisfies:
\eqn\rhor{|R_h(P)|=\left|{\GG({1\over 2}+iP)\over\GG({1\over 2})}\right|^2
={1\over \cosh(\pi P)}~.}
Hence,
\eqn\rsone{|R_h(P)|<1~,}
namely, part of the incoming wave is reflected from the black hole horizon.  

On the other hand, the scattering of winding modes on the cigar geometry
has a different interpretation \giv. It is analytically continued
to the scattering of momentum modes incoming from the asymptotically flat
regime behind the black hole singularity, and scattered from the singularity.
The result \rsp\ shows that the NS-NS scalars are fully reflected from 
the singularity:
\eqn\rsin{|R_s(P)|=1~.}
All in all, 
the results in this subsection show that the scattering of NS-NS scalars
in the $2$-d type~0 black hole is rather similar to the scattering in the 
bosonic string theory on the $2$-d black hole~\foot{Though
the 2-p-f in the bosonic and type 0 cases are {\it different} 
in a significant way, which will be important later.}
\refs{\dvv,\grs}, 
where~\foot{The exact two point function on the sphere
in the 2-d bosonic black hole
is obtained by eq. \sltpf\ with $\k=\k_B-2={1\over 4}$, 
$j=-{1\over 2}+i{P\over 2}$, and the on shell condition 
$m=\mp\bar m=i{3P\over 2}$. References \refs{\dvv,\grs} do not have the
$\k$-dependent phase factor in \sltpf, since they analyze only the 
classical limit (equivalent to $\k\to\infty$ in \sltpf).}
\eqn\rbos
{\eqalign{Bosonic \quad {SL(2)_{k_B=9/4}\over U(1)}:\quad &
j=-{1\over 2}+i{P\over 2}~,\cr
m=-\bar m=i{3P\over 2}~:\quad & 
R_h(P)={\GG(1-4iP)\GG(-iP)\over\GG(1+4iP)\GG(iP)}
\left({\GG({1\over 2}+2iP)\over\GG({1\over 2}-iP)}\right)^2~,\cr
& |R_h(P)|={\cosh(\pi P)\over\cosh(2\pi P)}~,\cr
m=\bar m=i{3P\over 2}~:\quad &
R_s(P)={\GG(1-4iP)\GG(-iP)\GG({1\over 2}+2iP)\GG({1\over 2}-iP)\over
        \GG(1+4iP)\GG(iP)\GG({1\over 2}-2iP)\GG({1\over 2}+iP)}~,\cr
&|R_s(P)|=1~.
}}
In the next subsection we discuss the scattering of R-R fields. 

\subsec{Scattering of R-R Fields}

In the type 0B string theory 
the R-R fields have $m=\bar m$ \mmrr, and thus
they carry only winding numbers \kkw\ on the cigar.
Thus the result \rzero\ indicates that these R-R fields are 
fully transmitted through the 
singularity~\foot{Wave
functions scattered behind the singularity of bosonic 2-d black holes 
were considered recently in \grs. In the charged cases it was found that
such wave functions are partly transmitted through the singularity.
However, in the uncharged case the wave functions are fully reflected from
the singularity \refs{\dvv,\grs}. 
On the other hand, in type 0B we find that, 
unlike the bosonic case and the NS-NS scalars in type~0, the R-R scalars are 
not fully reflected from the singularity. 
Instead, they are {\it fully transmitted} through the singularity.}.
This result is puzzling, since $R_s(P)=0$ both for the dispersion
relation $E=P=-ik$ (where 
$E=-i(m+n+\bar m +\bar n)=-2i(m+{1\over 2})$ 
is the energy and $P=-2i(j+1/2)$ is the momentum),
and for $E=-P=-ik$ (see appendix C). Perhaps it indicates that these R-R
scalars are non-propagating fields in the 2-d black hole background.

On the other hand, in the type 0A black hole, 
after a formal analytic continuation we found two possibilities.
The first is a propagating R-R
field (see \tststpf,\rzeroo),  with $m=-\bar m$ \mmmrr, 
which thus carries only momentum modes on the cigar \kkp. 
This scalar field is scattered from the horizon. 
In eq. \rzeroo, the dispersion relation is $E=-P$
($E=-i(m+n-\bar m -\bar n)=-2i(m+{1\over 2})=-ik$,
$P=-2i(j+{1\over 2})=ik$ in $\tilde\S^-(k)$; see appendix D). 
Hence, \rzeroo\ is the inverse of the reflection coefficient; 
it satisfies: 
\eqn\refcoa{|R_h(P)|^{-1}={\sinh(\pi P)\over \pi P}>1~,} 
(see $|$\rzeroo$|$).
The reflection coefficient itself is obtained for the correlator of scattering
waves with $E=P$ (see appendix D).
The other possibility (the analytic continuation of \tststpff) is 
$R_h(P)=0$. The latter is supported by classical space-time 
considerations~\foot{The R-R momentum modes should be described 
by a massless scalar which does not couple to the dilaton.
Since the massless Laplace equation in two space-time dimensions 
is conformally invariant (upon taking into account the regularity 
conditions on the horizon), we see that the reflection coefficient should 
vanish.}. 

%Finally, for $k>0$ ($E=P=-ik$) we found $R_h(P)=0$ \rzeroo,
%describing a scattering wave which is fully transmitted through the 
%horizon~\foot{We
%have not obtained the inverse processes for reasons that were discussed 
%in a couple of footnotes above. 
%The technical reason is that the transformation 
%which changes the relative sign of $P$ and $E$, 
%$(j,m,\bar m)\to (-(j+1),m,\bar m)$,
%takes an R-R physical operator to a non BRST invariant one
%(see appendix A)}. 
%{\bf xxx What is the physics of this???}
%Again, perhaps such a scalar is decoupled from the black hole 
%geometry as in the type 0B case~\foot{After T-duality, 
%which interchanges type 0B and type 0A, the results in this 
%subsection, concerning which are the propagating R-R fields in the 2-d 
%fermionic string, can be compared to \six.}. 

\newsec{A Matrix Model Dual -- A Conjecture}

In this section we speculate on an open/closed string duality between 
the type 0 $SL(2)_{1\over 2}/U(1)$ black hole, and the large
$N$ decoupled quantum matrix model on its (unstable) D-branes. 
Our conjecture follows a similar conjecture between the type 0 string
theory on Super-Liouville~$\times~\hat c=1$ matter \refs{\tt,\six}.
Hence, we first review the main essence of the conjecture of
\refs{\tt,\six}.

First, it is believed \refs{\mv,\kms,\mtv} that the bosonic closed string
theory on Liouville~$\times~c=1$ matter is dual to the 
matrix quantum mechanics
on the decoupled theory of its ZZ (1,1) branes \zz. 
The ZZ branes are D-branes localized 
in the strong coupling regime of the Liouville scalar $\phi$ 
(which has a linear dilaton). Their back reaction is expected to turn on 
the Liouville potential.
The decoupled quantum mechanics in the large $N$ limit of such D-branes   
is conjectured \refs{\mv,\kms,\mtv}
to be the ``old'' matrix model 
(for a review, see \refs{\klebanov,\gm})
of the 2-d bosonic string.

The ZZ branes are not stable. The open string tachyon potential has a local 
maximum. One side of this maximum is expected to have a local minimum
while the other side is not bounded from below 
(see \sen\ and references therein).
Hence, the matrix model dual is unstable non-perturbatively.
This is reflected in the fact that the bosonic matrix eigenvalues act as
free fermions in an inverted harmonic oscillator potential, 
which fill only one side of the potential up to a certain Fermi sea level. 
This potential is related to the open string tachyon potential, which turns
in the double scaling limit (leading to the large $N$ matrix model) into the
inverted harmonic oscillator potential.

A matrix model whose eigenvalues act as free fermions which fill the
inverted harmonic oscillator potential symmetrically on both sides is
conjectured \refs{\tt,\six} to be the open string dual of the type 0B 
closed string theory on Super-Liouville~$\times~\hat c=1$ matter.
Now, the inverted harmonic oscillator potential is associated with
the open string tachyon potential of unstable D-branes in the two dimensional 
type 0 fermionic string. This potential is symmetric around its local maximum
and has, in particular, local minima on both sides
(see \sen\ and references therein).
Hence, in the double scaling limit one obtains a stable vacuum
where the eigenvalues fill the two sides of the potential to the same Fermi
sea level.

A main ingredient in checking this duality is a simple relation between
correlators in the 2-d type 0 theory and the 2-d bosonic string.
An intuitive reason to this similarity is revealed by inspecting the
Liouville potential and the on-shell vertex operators in both cases.
In the 2-d bosonic string theory the Liouville potential is:
\eqn\lbos{\LL_B'=\mu_Be^{-\sqrt{2}\phi}=\mu_Be^{-{Q_B\over 2}\phi}~, 
\qquad Q_B=2\sqrt{2}~,}
where $\phi$ is the Liouville scalar field with a background charge $Q_B$.
On-shell vertex operators are:
\eqn\tbos{T^{\pm}_B(k)=e^{ikx}e^{(-{Q_B\over 2}\pm k)\phi}~,\qquad
Q_B=2\sqrt{2}~,}
where $x$ is the $c=1$ scalar.

On the other hand, in the 2-d type 0 theory the Super-Liouville potential is:
\eqn\lfer{\LL'=\mu\int d^2\theta e^{-\Phi}
=\mu\int d^2\theta e^{-{Q\over 2}\Phi}~, 
\qquad Q=2~,}
where $\Phi$ is a scalar superfield with a background charge $Q$,
whose physical components are $\phi$ and $\psi_\phi$.
The on-shell vertex operators (in the NS sector) are:
\eqn\tfer{T^{\pm}_{NS}(k)=e^{-\varphi}e^{ikx}e^{(-{Q\over 2}\pm k)\phi}~,
\qquad Q=2~.}
The  
similarity between \lbos,\tbos\ and \lfer,\tfer, respectively,
leads \dfk\ to correlators in the type 0 theory which are closely related to 
those in the bosonic theory: they are the same, up to leg factors and 
a rescaling $P\to \sqrt{2}P_B$ (the $\sqrt{2}$ originating from the 
ratio ${Q_B\over Q}=\sqrt{2}$).
 
Another 2-d bosonic, closed string theory which has a matrix model dual
is the bosonic string on sine-Liouville \kkk. 
Sine-Liouville is a theory of a real scalar $\phi$ with a background 
charge $Q_B$, a compact scalar $x$ on a circle with radius 
$R$, and a potential:~\foot{We choose $\alpha'=1$, hence we normalize 
$R$ as in \kkk, such that
$(k,\bar k)={1\over\sqrt{2}}({p\over R}+wR,{p\over R}-wR)$,
though unlike \kkk\ we work with canonically normalized scalars 
$\phi$ and $x$, as in \metas. The self-dual radius is 
${R\over\sqrt{\alpha'}}=1$, though
the $R\to{\alpha'\over R}$ T-duality is broken in the presence of the winding
mode condensate (the sine-Liouville potential).} 
\eqn\lsl{\LL'_{sl}=\lambda_{sl}e^{{R-2\over\sqrt{2}}\phi}
e^{i{R\over\sqrt{2}}(x(z)-\bar x(\bar z))}+c.c~.}  
As in Liouville~$\times~c=1$ matter, 
the Virasoro central charge of sine-Liouville is
\eqn\csl{c_{sl}=2+3Q_B^2~,}
and hence, criticality of the bosonic string implies
\eqn\cslqb{c_{sl}=26\quad\Leftrightarrow\quad Q_B=2\sqrt{2}~.}
In this case, $\LL'_{sl}$ in \lsl\ is marginal:
\eqn\truemarg{h(\LL'_{sl})={R^2\over 4}-{1\over 2}{R-2\over\sqrt{2}}
\left({R-2\over\sqrt{2}}+2\sqrt{2}\right)=1~.}
The matrix model dual of this string theory is studied in \kkk.
Again, the matrix eigenvalues act as free fermions in an inverted 
harmonic oscillator potential, but this time the eigenvalues dynamics 
involves not only the scalar representation of $SU(N)$, but also higher 
representations. In any case, as in the ``old'' matrix model, discussed above,
the eigenvalues fill only one side of the potential.

We finally return to the fermionic string on $SL(2)\over U(1)$.
The SCFT on the $SL(2)_{\k}\over U(1)$ cigar is T-dual to 
the $N=2$ Liouville theory \refs{\gk,\gktwo,\hk,\tong}.
The $N=2$ Liouville theory has a scalar $\phi$ with a background charge $Q$,
a compact scalar $x$:
\eqn\rtaq{x\simeq x+2\pi r~, \qquad r={\sqrt{2\alpha'}\over Q}~,}
and a superpotential:
\eqn\lntwol{\LL'_{N=2}=\lambda\int d^2\theta e^{-{1\over Q}(\Phi+i\tilde X)}
+c.c~, \qquad \tilde X\equiv \bar X(\bar z)-X(z)~.}
Here $\Phi$ is the superfield with physical components $\phi$ and 
$\psi_\phi$, and $X$ is a scalar superfield with physical components
$x$ and $\psi_x$, where $\phi$ and $x$ are given in \metas,\rphiq, and
$Q$ and $\k$ are related as in \rphiq. As before, in a critical fermionic 
string on $N=2$ Liouville:
\eqn\cqkntwo{c_{N=2}=15\quad\Leftrightarrow\quad Q=2\quad (\k={1\over 2})~.}
In this case, 
\eqn\lqtwo{\LL'_{N=2}(Q=2)=\lambda\int d^2\theta 
e^{-{1\over 2}(\Phi+i\tilde X)}
+c.c = \lambda\int d^2\theta e^{-{Q\over 4}(\Phi+i\tilde X)}+c.c~,}
where the compactification radius of $x$ is 
${r\over\sqrt{\alpha'}}={1\over\sqrt{2}}$ \rtaq\ (in which case
the $c=1$ CFT of $x$ is equivalent to the theory of a free 
``Dirac fermion'').

Following similar ideas to those in \refs{\tt,\six}, discussed above,
we are led to conjecture that  
the matrix model dual of type 0B string theory
on $N=2$ Liouville is the symmetric potential version of the KKK matrix model
\kkk\ at the self-dual radius
${R\over\sqrt{\alpha'}}=1$. 
Its non-perturbatively stable vacuum consists of matrix 
eigenvalues acting as free fermions, which
fill both sides of the potential in a symmetric way.
The reason that we have picked up the ${R\over\sqrt{\alpha'}}=1$ 
theory is because at this value
of $R$ the sine-Liouville potential \lsl\ is:
\eqn\lsls{\LL'_{sl}=\lambda_{sl}e^{-{\sqrt{2}\over 2}(\phi+i\tilde x)}+c.c.
=\lambda_{sl}e^{-{Q_B\over 4}(\phi+i\tilde x)}+c.c~,\qquad 
\tilde x\equiv \bar x(\bar z)-x(z)~, \quad Q_B=2\sqrt{2}~.}  
Moreover, the asymptotic behavior of the on-shell vertex operators in 
the weak coupling regime of sine-Liouville and $N=2$ Liouville is the 
same as \tbos\ and \tfer, respectively.
Comparing \lsls\ to \lqtwo\ we see that the bosonic string theory on
the ${R\over\sqrt{\alpha'}}=1$ sine-Liouville theory
has similar correlators to those of the type 0B string on $N=2$ Liouville
(up to leg factors and $P\to \sqrt{2}P_B$, originated from the ratio:
${Q_B\over Q}=\sqrt{2}$).

Since the $N=2$ Liouville is T-dual to the SCFT on $SL(2)\over U(1)$,
we are finally led to our main conjecture:
{\it the type 0A string theory on the $SL(2)_{1\over 2}/U(1)$ black hole
is dual to the ${R\over\sqrt{\alpha'}}=1$ KKK matrix model 
\kkk\ with a symmetric potential
and eigenvalues on both sides of its maximum.}

\vskip .1in
\noindent
A few comments are in order:

\item{(a)} 
We further conjecture that this dual matrix model is the decoupled
theory on $N\to\infty$ D-branes localized 
near the tip of the $SL(2)\over U(1)$ cigar. 
These are the analogs of the ZZ branes discussed above \refs{\gks,\pst}.
It is clear that they must exist for various reasons. One is that there are
consistent configurations of D-branes stretched between NS5-branes
(for a review, see \gkrev), and the 
near horizon limit of certain distributions of NS5-branes gives the 
$SL(2)\over U(1)$ SCFT \refs{\gk,\gktwo}. A D-brane stretched between
two such NS5-branes must turn into a D-brane localized near the tip of 
the cigar.
Another reason is that localized D-branes in~\foot{Actually,
in $H_3^+$ -- the Euclidean continuation of $SL(2)$.} 
$AdS_3\simeq \tilde{SL}(2)$ 
were found in 
\refs{\gks,\pst}, following a similar route to ZZ \zz. 
By taking the ``Fourier transform'' (as in \refs{\gk,\gktwo})
of the closed string 1-p-f on the disc, computed in
\refs{\gks,\pst}, and inducing 
the result to the $SL(2)\over U(1)$ quotient CFT,
we find the analog of the ZZ brane construction in the cigar CFT.
%{\bf xxx Shall we present these 1-p-f in an appendix ???}

\item{(b)} 
The sine-Liouville theory at the self-dual radius
${R\over\sqrt{\alpha'}}=1$ is not the one dual to the 
$SL(2)_{9\over 4}/U(1)$ bosonic 2-d black hole.
The latter is conjectured to be dual to sine-Liouville at 
${R\over\sqrt{\alpha'}}={3\over 2}$ \kkk.
One can see that the matrix model dual to type 0 on the 2-d black hole cannot
be the symmetric version of the KKK matrix model dual to the 2-d bosonic black
hole by inspecting eq. \rbos: the on-shell condition in the bosonic case
is $m=\pm\bar m=i{3\over 2}P$ while the on-shell condition in type 0 is 
$m=\pm\bar m=i{P\over 2}$
(the $SL(2)$ levels are also different:
$\k={1\over 2}$ in type 0 versus 
$\k\equiv\k_B-2={1\over 4}$ in the bosonic case). 
These lead to rather different correlators \sltpf\ in 
$SL(2)\over U(1)$ (for instance, compare \rbos\ to \rhp,\ttkttk\ and \rhor).  

\item{(c)}
The ${R\over\sqrt{\alpha'}}=1$ point (the self-dual radius)
in the investigation of \kkk\ is rather special. 
The methods used in \kkk\ break down precisely as 
${R\over\sqrt{\alpha'}}$ is decreased below $1$.
%~\foot{xxx {\bf Could
%it be that the $0$'s in the R-R correlators are related to the $0$'s appearing
%in the coordinate transformations used in \kkk\ at $R=1$?}}
Hence, it would be interesting to reconsider the matrix model at
this special point.

\item{(d)}
Our conjecture should of course be supported by more flesh, similar to \six.
For instance, it would be nice to prove the analog of eq. (3.47) in \six\
(the factorization of the correlators of certain linear combinations
$T_{L,R}$ of our ${\bf T}$ and ${\bf S}$),
to find the ground ring in the type 0 2-d black hole (or $N=2$ Liouville)
and the $\lambda$-dependent analog of eq. (3.39) in \six, and to
compute the torus partition function (though it would make sense to do it
in the models considered in (e), which depend on more parameters).  

\item{(e)}
It would be interesting to study the 2-d 
type 0 string theory on $R_{\phi}\times S_x^1$ \metas\
with a more generic superpotential: 
\eqn\llll{\int d^2\theta\left(\mu e^{-\Phi}+
\lambda e^{{R-2\over 2}\Phi-i{R\over 2}\tilde X)}+c.c.\right)~.}
It is likely that it is dual to a symmetric version
of the $(\mu,\lambda,R)$-dependent family of KKK matrix 
models~\foot{There is a difficulty here (which applies both to \llll\
and \lsl), pointed out to us by N. Seiberg: it is likely that
there are only 2 independent truly marginal deformations of the cigar CFT
(or its T-dual Liouville). The value of the asymptotic radius $R$ is 
correlated with $\lambda$ (as long as $\lambda\neq 0$).
Thus one should clarify what is the KKK \kkk\ string theory at 
${R\over\sqrt{\alpha'}}=1$.}. 

\bigskip
\noindent{\bf Acknowledgements:}
We thank N. Seiberg for discussions. 
This work is supported in part by the Israel 
Academy of Sciences and Humanities -- Centers of Excellence Program, 
the German-Israel Bi-National Science Foundation, the European RTN network
HPRN-CT-2000-00122, and the Horowitz foundation (AS).

\appendix{A}{BRST Invariance}

%{\bf xxx Should we add BRST invariance of the NS sector operators???}

In this appendix we will check the BRST invariance of the 
Ramond vertex operators in the $-{1\over 2}$ and $-{3\over 2}$ pictures.
%\sr,\sas.
Let us begin with the asymptotic regime, 
where the $N=2$ superconformal matter system is given by
\eqn\ltgj
{\eqalign{
T_m & = -\frac12 \d\phi\d\phi -\frac12 \psi_{\phi}\d\psi_{\phi} 
-\frac{Q}{2} \d^2\phi -\frac12 \d x\d x -\frac12 \psi_{x}\d\psi_{x}~,
\cr
G^{\pm} & = \frac{i}{2} (\psi_{\phi} \pm i \psi_{x}) 
\d ({\phi} \mp i x) + \frac{i}{2}Q \d (\psi_{\phi} \pm i \psi_{x})~, \cr
J^{U(1)} & = i \psi_{x} \psi_{\phi}  + i Q \d x~,\qquad\qquad Q=2~.
}}
The BRST operator can be written as 
\eqn\brsdec
{\eqalign{
Q_{BRST} = & Q_0 + Q_1 + Q_2~, \cr
Q_0 = & \oint c T(\phi, x,\psi_\phi,\psi_x,\beta, \gamma) + bc \d c~,\cr
Q_1 = & -\oint \gamma G_m = - \oint  e^{\varphi} \eta G_m~, \cr
Q_2= & - \frac14 \oint b \gamma^2 = - \frac14 \oint b e^{2 \varphi}\eta^2~,
}}
where the matter $N=1$ supercurrent $G_m$ is
\eqn\sc
{
G_m = G^+ + G^- =  i \psi_{\phi} \partial \phi +  
i \psi_{x} \d x + i Q \d \psi_{\phi}~,
}
and $T(\phi,x,\psi_\phi,\psi_x,\beta,\gamma)$ is the sum of $T_m$
and the stress-energy tensor for the $(\beta,\gamma)$ system of the
superconformal ghosts.
Note that we further bosonize the latter as $\beta=e^{-\varphi} \d \xi,
\gamma=e^{\varphi} \eta$, as usual,
with $\varphi(z) \varphi(w) \sim -\log(z-w)$ and the fermionic $(\xi, \eta)$
system of conformal dimension
$(0,1)$ satisfying $\xi \xi \sim \eta \eta \sim 0, \, \xi(z) \eta(w) \sim
\frac{1}{z-w}$.

The primaries of the Ramond sector in the SCFT \ltgj, 
which satisfy the mass shell 
condition $h={5\over 8}$ in the $-{1\over 2}$ and $-{3\over 2}$ pictures, 
are:
\eqn\rmv
{
{\cal S}^{\eps_1, \eps_2}(k) = 
e^{i{\eps_{2} \over 2}H' } e^{ikx}e^{(-1 + \eps_1 k)\phi}~, \qquad  
\eps_{1,2}= \pm 1~,
}
where $H'$ is given by \hprbos:
\eqn\bsn{
 \d H'  = \psi_x  \psi_{\phi} \,\,\,\,\,\, \Longleftrightarrow  \,\,\,\,\,\,
 e^{\pm i H'}  = {1\over \sqrt{2}}(\psi_{\phi} \pm i \psi_{x})~.
}
Using \bsn\ and $Q=2$, we can express
\eqn\scb
{
G_m= \frac{i}{\sqrt{2}} (e^{i H'} + e^{-i H'}) \partial \phi +  
\frac{1}{\sqrt{2}} (e^{i H'} - e^{-i H'}) \d x
+ i\sqrt{2} \d (e^{i H'} + e^{-i H'})~.
}
Now one obtains:
%\eqn\opegs{\eqalign{
%G(z) & S^{\eps_1, \eps_2}(k)(w) \sim  \cr
%& \frac{i}{\sqrt{2}} {(1- \eps_1 k) \over z-w}
%\left[ e^{i (\ed +1)H'}(z-w)^{\ed} + e^{i (\ed - 1)H'}(z-w)^{-\ed} \right] 
%e^{(-1 + \eps_1 k)\phi}  e^{ikx}  \cr
%& - \frac{i}{\sqrt{2}} {k \over z-w} \left[ e^{i (\ed +1)H'}(z-w)^{\ed} - 
%e^{i (\ed - 1)H'}(z-w)^{-\ed} \right] e^{(-1 + \eps_1 k)\phi}  e^{ikx}  \cr
%& + i \sqrt{2} \d_z \left[ e^{i (\ed +1)H'(w)}(z-w)^{\ed} + 
%e^{i (\ed - 1)H'(w)}(z-w)^{-\ed} \right] e^{(-1 + \eps_1 k)\phi}  e^{ikx} 
%\,\,\,.
%}}
\eqn\tdc{\eqalign{
&G_m(z){\cal S}^{\eps_1, \eps_2}(k)(w) \sim \cr
&- \frac{ik}{\sqrt{2}} 
e^{(-1 + \eps_1 k)\phi}  e^{ikx}
\left[ {(\eps_1 +1) \over (z-w)^{1 - \ed}} e^{i (\ed +1)H'} + 
{(\eps_1 -1) \over (z-w)^{1 + \ed}} e^{i (\ed -1)H'} \right]+...~,
}}
where the ``...'' stand for other terms with no $(z-w)^{-\frac32}$ 
factors.
Hence,
\eqn\gese
{
G_m(z){\cal S}^{\eps_1, \eps_2}(k)(w) \sim {-ik \over \sqrt{2}}
{\delta_{\eps_1, -\eps_2}{\cal S}^{\eps_1, -\eps_2}(k)(w) \over
(z-w)^{\frac32}}
+ {{\cal O} \over (z-w)^{\frac12}} + \cdots~.
}
We want to check whether $Q_{BRST}$ commutes with
\eqn\tvh{
V_{q}^{\eps_1, \eps_2}(k) \equiv e^{q\varphi}
{\cal S}^{\eps_1, \eps_2}(k)~, \qquad \e_{1,2}=\pm 1~, \qquad 
q=-{1\over 2}~, -{3\over 2}~.
}
{}From the form of $Q_1$ in \brsdec\ and \tvh,\gese, it follows
that in the $-{1\over 2}$ picture we should impose:
\eqn\brstinv{-{1\over 2}~~picture:\qquad \eps_1 = \eps_2~,}
for $k \neq 0$. On the other hand, in the
$-{3\over 2}$ picture all the four states \tvh\ are BRST closed.
However, the states in \tvh\ with $q=-{3\over 2}$,
$\eps_1=\eps_2$ and $k \neq 0$ are BRST exact, since they are given by
\eqn\bex
{
V_{-3/2}^{\eps_1, \eps_1}(k)(z)= \frac{i}{\sqrt{2}k} 
\left[Q_{BRST}, e^{-\frac{5
\varphi}{2}} \d \xi  {\cal S}^{\eps_1, -\eps_1}(k)(z) \right]~.
}
Note that  again the $Q_0$ and $Q_2$ terms do not contribute, and the action
of $Q_1$ gives the necessary simple pole.
As a final check of the BRST-exact character of 
$V_{-3/2}^{\eps_1,\eps_1}(k)$
and the non-triviality of 
$V_{-3/2}^{\eps_1, -\eps_1}(k)$,
we can apply the picture changing operator to both pairs of states,
obtaining:
\eqn\pco{\eqalign{
V_{-3/2}^{\eps_1, -\eps_1}(k) & \Longrightarrow \left[Q_{BRST},  \xi
V_{-3/2}^{\eps_1,
-\eps_1}(k) \right]
= -i\sqrt{2}k V_{-1/2}^{\eps_1, \eps_1}(k)~, \cr
V_{-3/2}^{\eps_1, \eps_1}(k) & \Longrightarrow \left[Q_{BRST},  \xi
V_{-3/2}^{\eps_1,
\eps_1}(k) \right] =0~.
}}
Thus we see that the picture-changing operator maps the BRST-exact states to
zero, and connects the non-trivial
states of the $-{3\over 2}$ and $-{1\over 2}$ pictures.
To summarize, in the $-{3\over 2}$ picture physical operators satisfy
\eqn\physop{-{3\over 2}~~picture: \qquad \e_1=-\e_2~,}
modulo BRST-exact operators with $\e_1=\e_2$. 
In the exact $SL(2)\over U(1)$ SCFT, the $N=2$ supercurrents and $R$-current 
of the superconformal algebra are given,
for instance, by the Kazama-Suzuki construction 
\kazsuz:~\foot{The $N=2$ $U(1)$ 
$R$-current $J_R$ was given in \jr\ (see \ih,\jx)~.}
\eqn\cuco{\eqalign{
G^{\pm} & = \frac1{\sqrt{\k}} \psi^{\pm}j^{\mp}~, \cr
J_R & =i\d H + {2\over\k} \, J^3 = i \sqrt{\frac{c}3} \d X_R~,
}}
where the $j^a$ and $J^a$, $a=\pm,3$, are the bosonic and total currents,
respectively, of the parent $SL(2)$ SCFT, 
and $\psi^a$ are the free fermions of $SL(2)$
(see \jjj,\jjpp,\norms,\ppm). Thus, to check BRST invariance 
one should compute the OPE between
\eqn\vre{
{\cal S}^{\eps_1, \eps_2}(k)=
V^{\eps_2 \frac12}_{-\frac12 +\eps_1 \frac{k}{2}, 
\frac{k}{2}-\eps_2 \frac12}~, \qquad \e_{1,2}=\pm~,
}
and the $N=1$ supercurrent
\eqn\gex{
G_m  = \frac1{\sqrt{\k}} \left[ \psi^{+}j^{-} + \psi^{-}j^{+} \right]~.
}
The result is the same as before ($\eps_1 = \eps_2$ in the $-\frac12$
picture). This can be seen in the following way.
One knows what is the action of $j^{\pm}$ and $\psi^{\pm}$ on
the $SL(2)$ primaries $\Phi_{jm}$ and $e^{iH}$ (see \bos), respectively.
Then using \dec, \expnh\ and the fact that 
$j^{\pm}={\tilde j}^{\pm} e^{\pm \sqrt{2 \over k+2}x_3}$ 
with ${\tilde j}^{\pm} x_3 \sim 0$ ($x_3$ is defined in \jxx),
one finds the OPE between the combinations $\psi^{\pm}j^{\mp}$ in $G_m$ 
and $V_{jm}^{n}$ \vjmn.
The result \brstinv\ then follows for the $e^{-{\varphi\over 2}}V_{jm}^{n}$  
given in \vre.
In the $-{3\over 2}$ picture all the four operators 
$e^{-{3\varphi\over 2}}V_{jm}^{n}$ in \vre\ are BRST-invariant
(though two independent linear combinations of them are 
exact~\foot{Those can be obtained by inspecting the asymptotic 
behavior mentioned in footnote 8 combined with \physop.}).

%Alternatively, one can bosonize the whole supersymmetric  
%$SL(2)\over U(1)$ coset, using the three free fields
%$H'$, $\phi$ and $x$, and show that \vre\ is given exactly by 
%\rmv, and that \gex\ is mapped into \scb.
%Then BRST invariance follows from the asymptotic result.
%{\bf xxx Either we add this construction or ERASE this paragraph ???}

\appendix{B}{2-P-F in the NS-NS Sector}

Consider first the case $k>0$ ($j=-{1\over 2}+{k\over 2}$) and $m=-\bar m$.
Inserting the appropriate values of $j,m,\bar m$ and $\k={1\over 2}$ 
into eq. \sltpf, and using eq. \tpf, we find that:~\foot{Although 
the unitarity bound \jreal\ actually does not allow $j={p-1\over 2}$, 
and even though for such values of $j$ there are
(double) poles in the Euclidean 2-p-f,
we write it formally as we are
interested in its analytic continuation below.}
\eqn\ttkp{\eqalign{
&k=p\in Z_+~,\quad m=-\bar m={p\over 2}: \cr 
&\langle \T^{--}(-p)\T^{+-}(p)\rangle=
\langle V_{-{1\over 2}+{p\over 2},-{p\over 2},{p\over 2}}
V_{-{1\over 2}+{p\over 2},{p\over 2},-{p\over 2}}\rangle=
\left({\GG(1-2p)\GG(-p)\over\GG(1+2p)\GG(p)}\right)
\left({\GG({1\over 2}+p)\over\GG({1\over 2})}\right)^2~.
}}
After analytic continuation $k=p\to iP$, $P\in R$, 
to the $1+1$ Lorentzian black hole,
we have:
\eqn\rofp{R_h(P)\equiv\langle \T^{--}(-iP)\T^{+-}(iP)\rangle=
\left({\GG(1-2iP)\GG(-iP)\over\GG(1+2iP)\GG(iP)}\right)
\left({\GG({1\over 2}+iP)\over\GG({1\over 2})}\right)^2~.}
Next, for $k>0$ and $m=\bar m$:
\eqn\ttkpw{\eqalign{
&k={w\over 2}\in {Z_+\over 2}~,\quad m=\bar m={w\over 4}: \cr 
&\langle \T^{-+}(-{w\over 2})\T^{++}({w\over 2})\rangle=
\langle V_{-{1\over 2}+{w\over 4},-{w\over 4},-{w\over 4}}
V_{-{1\over 2}+{w\over 4},{w\over 4},{w\over 4}}\rangle=
\left({\GG(1-w)\GG(-{w\over 2})\over\GG(1+w)\GG({w\over 2})}\right)
{\GG({1\over 2}+{w\over 2})\over\GG({1\over 2}-{w\over 2})}~.
}}
After analytic continuation $k={w\over 2}\to iP$, $P\in R$, 
to the Lorentzian black hole, we have:
\eqn\rofw{R_s(P)\equiv\langle \T^{-+}(-iP)\T^{++}(iP)\rangle=
\left({\GG(1-2iP)\GG(-iP)\over\GG(1+2iP)\GG(iP)}\right)
{\GG({1\over 2}+iP)\over\GG({1\over 2}-iP)}~.}
We have thus obtained the 2-p-f of positive momentum modes 
and positive winding modes on the Euclidean cigar, and their continuation
to the $1+1$ Lorentzian black hole. 
We now turn to negative momenta and windings
and their analytic continuation.

For $k<0$ ($j=-{1\over 2}-{k\over 2}$) and $m=-\bar m$ one finds:
\eqn\ttkm{\eqalign{
&k=p\in Z_-~,\quad m=-\bar m={p\over 2}: \cr 
&\langle \T^{+-}(-p)\T^{--}(p)\rangle=
\langle V_{-{1\over 2}-{p\over 2},-{p\over 2},{p\over 2}}
V_{-{1\over 2}-{p\over 2},{p\over 2},-{p\over 2}}\rangle=
\left({\GG(1+2p)\GG(p)\over\GG(1-2p)\GG(-p)}\right)
\left({\GG({1\over 2})\over\GG({1\over 2}+p)}\right)^2~.
}}
Its analytic continuation to the Lorentzian quotient gives:
\eqn\ttmmmm{\langle \T^{+-}(-iP)\T^{--}(iP)\rangle=R_h^{-1}(P)~,}
where $R_h(P)$ is given in \rofp.

Finally, for $k<0$ and $m=\bar m$:
\eqn\ttkmm{\eqalign{
&k={w\over 2}\in {Z_-\over 2}~,\quad m=\bar m={w\over 4}: \cr 
&\langle \T^{++}(-{w\over 2})\T^{-+}({w\over 2})\rangle=
\langle V_{-{1\over 2}-{w\over 4},-{w\over 4},-{w\over 4}}
V_{-{1\over 2}-{w\over 4},{w\over 4},{w\over 4}}\rangle=
\left({\GG(1+w)\GG({w\over 2})\over\GG(1-w)\GG(-{w\over 2})}\right)
{\GG({1\over 2}-{w\over 2})\over\GG({1\over 2}+{w\over 2})}~,
}}
and its analytic continuation gives:
\eqn\ttmmmm{\langle \T^{++}(-iP)\T^{-+}(iP)\rangle=R_s(-P)=R_s^{-1}(P)~,}
where $R_s(P)$ is given in \rofw.

\appendix{C}{2-P-F in the R-R Sector of Type 0B}

First of all, from eqs. \srr\ and \conserv\ we see that a non-vanishing
two point correlator in the R-R sector is between 
$\S_{-{1\over 2}}(k)\equiv \S(k)$ and (the ``dagger'' $\e_2\to -\e_2$ of) 
its image in the $(-{3\over 2},-{3\over 2})$ picture (see appendix A):
\eqn\srrp{\eqalign{
\S_{-{3\over 2}}^{\mp}(k)=&e^{-{3\varphi\over 2}-{3\bar\varphi\over 2}}
V_{-{1\over 2}\mp{k\over 2},{k\over 2}\mp{1\over 2},
{k\over 2}\mp{1\over 2}}^{\pm{1\over 2},\pm{1\over 2}}\cr 
=&e^{-{3\varphi\over 2}-{3\bar\varphi\over 2}}
V_{-{1\over 2}\mp{k\over 2},{k\over 2}\mp{1\over 2},{k\over 2}\mp{1\over 2}}
e^{{i\over\sqrt{5}}(2k\pm{1\over 2})(X_R(z)-\bar X_R(\bar z))}~.
}}
%Note that the only difference between $\S_{-{1\over 2}}(k)$ and 
%$\S_{-{3\over 2}}(k)$ is their $e^{q\varphi+q\bar\varphi}$ 
%dependence~\foot{This can be seen, for instance, by taking the on-shell 
%product of $\T_{-1}(k_1)\equiv\T(k_1)$ with 
%$\S_{-{1\over 2}}(k_2)\equiv \S(k_2)$, whose result is 
%$\S_{-{3\over 2}}(k_1+k_2)$.}.
Using the same methods as in the previous subsection, we find that
for $k>0$:
\eqn\sksk{\eqalign{
\langle \S_{-{3\over 2}}^-(-k)\S_{-{1\over 2}}^+(k)\rangle =&
\langle V_{-{1\over 2}+{k\over 2},{1\over 2}-{k\over 2},
{1\over 2}-{k\over 2}}
V_{-{1\over 2}+{k\over 2},-{1\over 2}+{k\over 2},-{1\over 2}+{k\over 2}}
\rangle \cr =&{\GG(1-2k)\GG(-k)\GG(k)\GG(1)\over
\GG(1+2k)\GG(k)\GG(1-k)\GG(0)}~.}}
%and hence
%~\foot{{\bf xxx Could it be that this means that $\S^+$ is null
%and hence BRST exact? Ari, can you check this?}} 
%\eqn\rzero{R_s(P)\equiv
%\langle\S_{-{3\over 2}}^-(-iP)\S_{-{1\over 2}}^+(iP)\rangle =0~.}
For $k<0$ we obtain:
\eqn\skskm{\eqalign{
\langle \S_{-{3\over 2}}^+(-k)\S_{-{1\over 2}}^-(k)\rangle =&
\langle V_{-{1\over 2}-{k\over 2},-{1\over 2}-{k\over 2},
-{1\over 2}-{k\over 2}}
V_{-{1\over 2}-{k\over 2},{1\over 2}+{k\over 2},{1\over 2}+{k\over 2}}
\rangle \cr =&{\GG(1+2k)\GG(k)\GG(1)\GG(-k)\over
\GG(1-2k)\GG(-k)\GG(0)\GG(1+k)}~.}}
%and hence~\foot{Note that here we get $R=0$ both for $k>0$ and $k<0$; howcome 
%we did not get also the inverse of the reflection coefficient -- 
%$\infty$ in this case? The answer is that in $SL(2)$ the correlators 
%which give this infinite inverse are:
%$\langle V_{-{1\over 2}+{k\over 2},-{1\over 2}-{k\over 2},
%-{1\over 2}-{k\over 2}}
%V_{-{1\over 2}+{k\over 2},{1\over 2}+{k\over 2},{1\over 2}+{k\over 2}}
%\rangle$
%for $k>0$, which give \skskm$^{-1}$,
%and
%$\langle V_{-{1\over 2}-{k\over 2},{1\over 2}-{k\over 2},
%{1\over 2}-{k\over 2}}
%V_{-{1\over 2}-{k\over 2},-{1\over 2}+{k\over 2},-{1\over 2}+{k\over 2}}
%\rangle$ 
%for $k<0$, which give \sksk$^{-1}$.
%However, R-R operators with such $V_{jm\bar m}$ are not BRST invariant
%(see appendix A).
%Technically, \skskm\ is not the inverse of \sksk\ because the two are 
%related by $(j;m,\bar m)\to (-(j+1),m+1,\bar m+1)$ (here we regard $j$ 
%as a free, unrestricted variable), while a transformation
%of $(j;m,\bar m)$ which gives the inverse is: 
%$(j;m,\bar m)\to (-(j+1),m,\bar m)$ (this is indeed the relation between
%\ttkp, \ttkpw\ and \ttkm, \ttkmm, respectively, in the NS-NS sector). 
%}
%\eqn\rinf{\langle\S_{-{3\over 2}}^+(-iP)\S_{-{1\over 2}}^-(iP)\rangle =
%R_s(-P)=0~.}
Note that since here $m=\bar m$, namely, $k={w\over 2}\in{Z\over 2}$,
we actually obtain that \sksk\ $=$ \skskm\ 
$= {2(-)^{w}\over (|w|!)^2}$.~\foot{Use $\GG(x+1)=x\GG(x)$ and
$\GG(-n+\e)={(-)^n\over \e n!}+O(1)$, $n=0,1,2,...$, to obtain this result.}
Due to the $\GG(0)$ in the denominator of eqs. \sksk,\skskm, it does not
make sense to analytically continue to Lorentzian space~\foot{One obtains
that in both cases the Lorentzian 2-p-f for the continuous series is
$R_s(P)=0$ for all $P\in R$ (note that the analytic continuation $k\to iP$
is done before we write $\GG(k)$ in terms of $k!$, which is valid
for $k\in Z$ but not for $k\in iR$).}.

Similarly, one finds that for $k>0$:
\eqn\skskmt{\eqalign{
\langle \S_{-{1\over 2}}^-(-k)\S_{-{3\over 2}}^+(k)\rangle =&
\langle V_{-{1\over 2}+{k\over 2},-{1\over 2}-{k\over 2},
-{1\over 2}-{k\over 2}}
V_{-{1\over 2}+{k\over 2},{1\over 2}+{k\over 2},{1\over 2}+{k\over 2}}
\rangle \cr 
=&\left({\GG(1+2k)\GG(k)\GG(1)\GG(-k)\over
\GG(1-2k)\GG(-k)\GG(0)\GG(1+k)}\right)^{-1}~,}}
and for $k<0$:
\eqn\skskt{\eqalign{
\langle \S_{-{1\over 2}}^+(-k)\S_{-{3\over 2}}^-(k)\rangle =&
\langle V_{-{1\over 2}-{k\over 2},{1\over 2}-{k\over 2},
{1\over 2}-{k\over 2}}
V_{-{1\over 2}-{k\over 2},-{1\over 2}+{k\over 2},-{1\over 2}+{k\over 2}}
\rangle \cr 
=&\left({\GG(1-2k)\GG(-k)\GG(k)\GG(1)\over
\GG(1+2k)\GG(k)\GG(1-k)\GG(0)}\right)^{-1}~.}}
Note that \skskmt\ and \skskt\ are the inverse of \skskm\ and \sksk, 
respectively. The comments below eq. \skskm\ thus apply here as well.

\appendix{D}{2-P-F in the R-R Sector of Type 0A}

We now have the two point correlator between
$\tilde \S_{-{1\over 2}}(k)\equiv \tilde \S(k)$ \srra\ 
and $\tilde \S_{-{3\over 2}}(k)$. The latter is the same as 
$\tilde \S(k)$ except for the changes 
$e^{-{\varphi\over 2}-{\bar\varphi\over 2}}\to
e^{-{3\varphi\over 2}-{3\bar\varphi\over 2}}$ (and $j\to -(j+1)$
if $\e_2=-\e_1$; see appendix A).

We find that for $k>0$:
\eqn\tsktsk{\eqalign{
\langle \tilde \S_{-{3\over 2}}^-(-k)\tilde \S_{-{1\over 2}}^+(k)\rangle =&
\langle V_{-{1\over 2}+{k\over 2},{1\over 2}-{k\over 2},
-{1\over 2}+{k\over 2}}
V_{-{1\over 2}+{k\over 2},-{1\over 2}+{k\over 2},{1\over 2}-{k\over 2}}
\rangle \cr =&{\GG(1-2k)\GG(-k)\over
\GG(1+2k)\GG(k)}\left(\GG(k)\over\GG(0)\right)^2~,}}
%and hence 
%\eqn\rzeroo{R_h(P)\equiv
%\langle\tilde\S_{-{3\over 2}}^-(-iP)
%\tilde\S_{-{1\over 2}}^+(iP)\rangle =0^2~.}
and for $k<0$ we obtain:
\eqn\tsktskm{\eqalign{
\langle \tilde\S_{-{3\over 2}}^+(-k)\tilde\S_{-{1\over 2}}^-(k)\rangle =&
\langle V_{-{1\over 2}-{k\over 2},-{1\over 2}-{k\over 2},
{1\over 2}+{k\over 2}}
V_{-{1\over 2}-{k\over 2},{1\over 2}+{k\over 2},-{1\over 2}-{k\over 2}}
\rangle \cr =&{\GG(1+2k)\GG(k)\over
\GG(1-2k)\GG(-k)}\left(\GG(1)\over\GG(1+k)\right)^2~.}}
%and hence~\foot{As before, the inverse of the correlators \tsktsk\ and 
%\tsktskm\ is obtained in $SL(2)$ as follows. When $k>0$:
%$\langle V_{-{1\over 2}+{k\over 2},-{1\over 2}-{k\over 2},
%{1\over 2}+{k\over 2}}
%V_{-{1\over 2}+{k\over 2},{1\over 2}+{k\over 2},-{1\over 2}-{k\over 2}}
%\rangle=$\tsktskm$^{-1}$, and when $k<0$:
%$\langle V_{-{1\over 2}-{k\over 2},{1\over 2}-{k\over 2},
%-{1\over 2}+{k\over 2}}
%V_{-{1\over 2}-{k\over 2},-{1\over 2}+{k\over 2},{1\over 2}-{k\over 2}}
%\rangle=$\tsktsk$^{-1}$.
%Again, R-R operators with such $V_{jm\bar m}$ are not BRST invariant
%(see appendix A). Moreover, the technical reason that \tsktskm\ is not
%the inverse of \tsktsk\ is as explained in the previous footnote for the
%R-R sector of type 0B (here \tsktskm\ is related to \tsktsk\ by
%$(j,m,\bar m)\to (-(j+1),m+1,\bar m -1)$).} 
%\eqn\rstrange{\langle\tilde\S_{-{3\over 2}}^+(-iP)
%\tilde\S_{-{1\over 2}}^-(iP)\rangle =
%{\GG(1+2iP)\GG(iP)\over
%\GG(1-2iP)\GG(-iP)}\left(\GG(1)\over\GG(1+iP)\right)^2~.}
Note that since here $m=-\bar m$, namely, $k=p\in Z$,
we actually obtain that \tsktsk\ $=$ \tsktskm\ 
$= {2(-)^{p-1}\over (|2p|!)^2}$.
The analytic continuation of \tsktskm\ gives 
(for $P<0$, since we choose the branch with $-i(j+{1\over 2})>0$):
\eqn\rstrange{\left(R_h(P)\right)^{-1}={\GG(1+2iP)\GG(iP)\over
\GG(1-2iP)\GG(-iP)}\left(\GG(1)\over\GG(1+iP)\right)^2~.}
Similarly, one finds that for $k>0$:
\eqn\skskmtt{\eqalign{
\langle \tilde\S_{-{1\over 2}}^-(-k)\tilde\S_{-{3\over 2}}^+(k)\rangle =&
\langle V_{-{1\over 2}+{k\over 2},-{1\over 2}-{k\over 2},
{1\over 2}+{k\over 2}}
V_{-{1\over 2}+{k\over 2},{1\over 2}+{k\over 2},-{1\over 2}-{k\over 2}}
\rangle \cr 
=&\left({\GG(1+2k)\GG(k)\over
\GG(1-2k)\GG(-k)}\right)^{-1}\left(\GG(1)\over\GG(1+k)\right)^{-2}~,}}
and for $k<0$:
\eqn\sksktt{\eqalign{
\langle \tilde\S_{-{1\over 2}}^+(-k)\tilde\S_{-{3\over 2}}^-(k)\rangle =&
\langle V_{-{1\over 2}-{k\over 2},{1\over 2}-{k\over 2},
-{1\over 2}+{k\over 2}}
V_{-{1\over 2}-{k\over 2},-{1\over 2}+{k\over 2},{1\over 2}-{k\over 2}}
\rangle \cr 
=&\left({\GG(1-2k)\GG(-k)\over
\GG(1+2k)\GG(k)}\right)^{-1}\left({\GG(k)\over\GG(0)}\right)^{-2}~.}}
Note that \skskmtt\ and \sksktt\ are the inverse of \tsktskm\ and \tsktsk, 
respectively. In particular, the analytic continuation of \skskmtt\
is $R_h(P)$ (for $P>0$,
since we choose the branch with $-i(j+{1\over 2})>0$), 
the inverse of \rstrange.

\listrefs
\end